\documentclass[12pt]{article}
\usepackage{axodraw}

\def\comp{{\rm C}\llap{\vrule height7.1pt width1pt depth-.4pt\phantom t}}
\def\ltwid{\mathrel{\raise.3ex\hbox{$<$\kern-.75em\lower1ex\hbox{$\sim$}}}}
\def \be{\begin{equation}}
\def \ee{\end{equation}}
\def \bea{\begin{eqnarray}}
\def \eea{\end{eqnarray}}

\def \del{\partial}
\def \a{\alpha}

\def \f{\frac}

\def \g{\gamma}
\def \nn{\nonumber}

\def \Om{\Omega}

\def \si{\sigma}

\def \d{\delta}
\def \D{\Delta}
\def \e{\eta}

\begin{document}

\begin{titlepage}

\begin{flushright}
gr-qc/0603135 \\ UFIFT-QG-06-01
\end{flushright}

\vspace{2cm}

\begin{center}
{\bf Gravitons Enhance Fermions during Inflation}
\end{center}

\vspace{.5cm}

\begin{center}
S. P. Miao$^{\dagger}$ and R. P. Woodard$^{\ddagger}$
\end{center}

\vspace{.5cm}

\begin{center}
\it{Department of Physics \\
University of Florida \\
Gainesville, FL 32611}
\end{center}

\vspace{1cm}

\begin{center}
ABSTRACT
\end{center}
We solve the effective Dirac equation for massless fermions during
inflation in the simplest gauge, including all one loop
corrections from quantum gravity. At late times the result for a
spatial plane wave behaves as if the classical solution were
subjected to a time dependent field strength renormalization of
$Z_2(t) = 1 - \frac{17}{4 \pi} G H^2 \ln(a) + O(G^2)$. We show that
this also follows from making the Hartree approximation, although
the numerical coefficients differ.

\vspace{2cm}

\begin{flushleft}
\end{flushleft}

\begin{flushleft}
$^{\dagger}$ e-mail: miao@phys.ufl.edu \\
$^{\ddagger}$ e-mail: woodard@phys.ufl.edu
\end{flushleft}

\end{titlepage}

\section{Introduction}

Gravitons and massless, minimally coupled scalars can mediate vastly
enhanced quantum effects during inflation because they are simultaneously
massless and not conformally invariant \cite{RPW1}. One naturally wonders
how interactions with these quanta affect themselves and other particles.
The first step in answering this question on the linearized level is to
compute the one particle irreducible (1PI) 2-point function for the field
whose behavior is in question. This has been done at one loop order for
gravitons in pure quantum gravity \cite{TW1}, for photons \cite{PTW1,PTW2}
and charged scalars \cite{KW} in scalar quantum electrodynamics (SQED), for
fermions \cite{PW1,GP} and Yukawa scalars \cite{DW} in Yukawa theory, for
fermions in Dirac + Einstein \cite{MW1} and, at two loop order, for scalars
in $\phi^4$ theory \cite{BOW}. The next step is using the 1PI 2-point
function to correct the linearized equation of motion for the field in
question. That is what we shall do here for the fermions of massless Dirac +
Einstein.

It is worth reviewing the conventions used in computing the fermion
self-energy \cite{MW1}. We worked on de Sitter background in conformal
coordinates,
\begin{equation}
ds^2 = a^2(\eta) \Bigl(-d\eta^2 + d\vec{x} \!\cdot\! d\vec{x}\Bigr) \qquad
{\rm where} \qquad a(\eta) = -\frac1{H \eta} = e^{H t} \; .
\end{equation}
We used dimensional regularization and obtained the self-energy for the
conformally re-scaled fermion field,
\begin{equation}
\Psi(x) \equiv a^{(\frac{D-1}2)} \psi(x) \; .
\end{equation}
The local Lorentz gauge was fixed to allow an algebraic expression for the
vierbein in terms of the metric \cite{RPW2}. The general coordinate gauge
was fixed to make the tensor structure of the graviton propagator decouple
from its spacetime dependence \cite{TW2,RPW3}. The result we obtained is,
\begin{eqnarray}
\lefteqn{\Bigl[\Sigma^{\rm ren}\Bigr](x;x') \!=\!\frac{i \kappa^2
H^2}{2^6 \pi^2} \Biggl\{\frac{\ln(a a')}{H^2 a a'} \hspace{-.1cm}
\not{\hspace{-.1cm}
\partial} \partial^2 \!+\! \frac{15}2 \ln(a a') \hspace{-.1cm} \not{\hspace{
-.1cm} \partial} \!-\! 7 \ln(a a') \; \hspace{-.1cm}
\overline{\not{\hspace{
-.1cm} \partial}} \Biggr\} \delta^4(x \!-\! x') } \nonumber \\
& & \hspace{.2cm} + \frac{\kappa^2}{2^8 \pi^4} (a a')^{-1}
\hspace{-.1cm} \not{\hspace{-.1cm} \partial} \partial^4 \Bigl[
\frac{\ln(\mu^2 \Delta x^2)}{ \Delta x^2} \Bigr] + \frac{\kappa^2
H^2}{2^8 \pi^4} \Biggl\{\Bigl(\frac{15}2 \hspace{-.1cm}
\not{\hspace{-.1cm} \partial} \, \partial^2 - \hspace{-.1cm}
\overline{\not{\hspace{-.1cm} \partial}} \, \partial^2 \Bigr)
\Bigl[ \frac{\ln(\mu^2 \Delta x^2)}{\Delta x^2} \Bigr] \nonumber \\
& & \hspace{1.5cm} + \Bigl(-8 \; \hspace{-.1cm}
\overline{\not{\hspace{-.1cm}
\partial}} \partial^2 \!+\! 4 \hspace{-.1cm} \not{\hspace{-.1cm} \partial}
\nabla^2 \Bigr) \Bigl[ \frac{\ln(\frac14 H^2\Delta x^2)}{\Delta x^2}
\Bigr] \!+\! 7 \hspace{-.1cm} \not{\hspace{-.1cm} \partial} \,
\nabla^2 \Bigl[ \frac1{\Delta x^2} \Bigr]\!\Biggr\} + O(\kappa^4) \; ,
\qquad \label{ren}
\end{eqnarray}
where $\kappa^2 \equiv 16 \pi G$ is the loop counting parameter of quantum
gravity. The various differential and spinor-differential operators are,
\begin{equation}
\partial^2 \equiv \eta^{\mu\nu} \partial_{\mu} \partial_{\nu} \;\; , \;\;
\nabla^2 \equiv \partial_i \partial_i \;\; , \;\; \hspace{-.1cm}
\not{\hspace{-.1cm} \partial} \equiv \gamma^{\mu} \partial_{\mu} \;\; {\rm and}
\;\; \hspace{-.1cm} \overline{\not{\hspace{-.1cm} \partial}} \, \equiv \gamma^i 
\partial_i \; ,
\end{equation}
where $\eta^{\mu\nu}$ is the Lorentz metric and $\gamma^{\mu}$ are the gamma
matrices. The conformal coordinate interval is basically $\Delta x^2 \equiv
(x\!-\!x')^{\mu} (x\!-\!x')^{\nu} \eta_{\mu\nu}$, up to a subtlety about the
imaginary part which will be explained shortly.

The linearized, effective Dirac equation we will solve is,
\begin{equation}
i\hspace{-.1cm}\not{\hspace{-.08cm} \partial}_{ij} \Psi_{j}(x) -
\int d^4x' \, \Bigl[\mbox{}_i \Sigma_j \Bigr](x;x') \, \Psi_{j}(x')
= 0 \; . \label{Diraceqn}
\end{equation}
In judging the validity of this exercise it is important to answer five
questions:
\begin{enumerate}
\item{What is the relation between the $\comp$-number, effective field
equation (\ref{Diraceqn}) and the Heisenberg operator equations of
Dirac + Einstein?}
\item{How do solutions to (\ref{Diraceqn}) change when different gauges
are used?}
\item{How do solutions to (\ref{Diraceqn}) depend upon the finite parts
of counterterms?}
\item{What is the imaginary part of $\Delta x^2$? and}
\item{What can we do without the higher loop contributions to the
fermion self-energy?}
\end{enumerate}
Issues 1 and 2 are closely related, and require a lengthy digression that we 
have consigned to section 2 of this paper. In this Introduction we will 
comment on issues 3-5.

Dirac + Einstein is not perturbatively renormalizable \cite{DVN}, so we 
could only obtain a finite result by absorbing divergences in the BPHZ sense 
\cite{BP,H,Z1,Z2} using three higher derivative counterterms,
\begin{equation}
-\kappa^2 H^2 \Bigl\{\frac{\alpha_1}{H^2 a a'} \hspace{-.1cm}
\not{\hspace{-.1cm} \partial} \partial^2 + \alpha_2 D (D\!-\!1) \hspace{-.1cm}
\not{\hspace{-.1cm} \partial} + \alpha_3 \; \hspace{-.1cm} \overline{\not{
\hspace{-.1cm} \partial}} \Bigr\} \delta^D(x\!-\!x') \; . \label{genctm}
\end{equation}
No physical principle seems to fix the finite parts of these counterterms so
any result which derives from their values is arbitrary. We chose to null
local terms at the beginning of inflation ($a = 1$), but any other choice
could have been made and would have affected the solution to (\ref{Diraceqn}).
Hence there is no point in solving the equation exactly. However, each of
the three counterterms is related to a term in (\ref{ren}) which carries a
factor of $\ln(a a')$,
\begin{eqnarray}
\frac{\alpha_1}{H^2 a a'} \hspace{-.1cm} \not{\hspace{-.1cm} \partial}
\partial^2 & \Longleftrightarrow & \frac{\ln(a a')}{H^2 a a'} \hspace{-.1cm}
\not{\hspace{-.1cm} \partial} \partial^2 \; , \label{1stlog} \\
\alpha_2 D (D\!-\!1) \hspace{-.1cm} \not{\hspace{-.1cm} \partial}
& \Longleftrightarrow & \frac{15}2 \ln(a a') \hspace{-.1cm} \not{\hspace{-.1cm}
\partial} \; , \label{2ndlog} \\
\alpha_3 \; \hspace{-.1cm} \overline{\not{\hspace{-.1cm} \partial}}
& \Longleftrightarrow & -7 \ln(a a') \; \hspace{-.1cm} \overline{\not{\hspace{-
.1cm} \partial}} \; . \label{3rdlog}
\end{eqnarray}
Unlike the $\alpha_i$'s, the numerical coefficients of the right hand terms
are uniquely fixed and completely independent of renormalization. The factors
of $\ln(a a')$ on these right hand terms mean that they dominate over any
finite change in the $\alpha_i$'s at late times. It is in this late time
regime that we can make reliable predictions about the effect of quantum
gravitational corrections.

The analysis we have just made is a standard feature of low energy effective
field theory, and has many distinguished antecedents
\cite{BN,SW,FS,HS,CDH,CD,DMC1,DL,JFD1,JFD2,MV,HL,ABS,KK1,KK2}. Loops of
massless particles make finite, nonanalytic contributions which cannot be
changed by counterterms and which dominate the far infrared. Further, these
effects must occur as well, with precisely the same numerical values, in
whatever fundamental theory ultimately resolves the ultraviolet problems
of quantum gravity.

We must also clarify what is meant by the conformal coordinate interval
$\Delta x^2(x;x')$ which appears in (\ref{ren}). The in-out effective field
equations correspond to the replacement,
\begin{equation}
\Delta x^2(x;x') \longrightarrow \Delta x^2_{\scriptscriptstyle ++}(x;x')
\equiv \Vert \vec{x} - \vec{x}' \Vert^2 - (\mid\e-\e'\mid-i\d)^2 \; .
\label{D++}
\end{equation}
These equations govern the evolution of quantum fields under the assumption
that the universe begins in free vacuum at asymptotically early times and ends
up the same way at asymptotically late times. This is valid for scattering in
flat space but not for cosmological settings in which particle production
prevents the in vacuum from evolving to the out vacuum. Persisting with the
in-out effective field equations would result in quantum correction terms
which are dominated by events from the infinite future! This is the correct
answer to the question being asked, which is, ``what must the field be in order
to make the universe to evolve from in vacuum to out vacuum?'' However, that
question is not very relevant to any observation we can make.

A more realistic question is, ``what happens when the universe is released
from a prepared state at some finite time and allowed to evolve as it will?''
This sort of question can be answered using the Schwinger-Keldysh formalism
\cite{JS,KTM,BM,LVK,CSHY,RDJ,CH}. For a recent derivation in the position-space
formalism we are using, see \cite{FW}. We confine ourselves here to noting four
simple rules:
\begin{itemize}
\item{The endpoints of lines in the Schwinger-Keldysh formalism carry a 
$\pm$ polarity, so every n-point 1PI function of the in-out formalism gives
rise to $2^n$ 1PI functions in the Schwinger-Keldysh formalism;}
\item{The linearized effective Dirac equation of the Schwinger-Keldysh
formalism takes the form (\ref{Diraceqn}) with the replacement,
\begin{equation}
\Bigl[\mbox{}_i \Sigma_j\Bigr](x;x') \longrightarrow 
\Bigl[\mbox{}_i \Sigma_j\Bigr]_{\scriptscriptstyle ++}\!\!\!\!(x;x') +
\Bigl[\mbox{}_i \Sigma_j\Bigr]_{\scriptscriptstyle +-}\!\!\!\!(x;x') \; ;
\end{equation}}
\item{The ${\scriptscriptstyle ++}$ fermion self-energy is (\ref{ren}) with 
the replacement (\ref{D++}); and}
\item{The ${\scriptscriptstyle +-}$ fermion self-energy is,
\begin{eqnarray}
\lefteqn{
- \frac{\kappa^2}{2^8 \pi^4 a a'} \hspace{-.1cm} \not{\hspace{-.1cm} \partial} 
\partial^4 \Bigl[\frac{\ln(\mu^2 \Delta x^2)}{ \Delta x^2} \Bigr] -
\frac{\kappa^2 H^2}{2^8 \pi^4} \Biggl\{\Bigl(\frac{15}2 \hspace{-.1cm}
\not{\hspace{-.1cm} \partial} \, \partial^2 - \hspace{-.1cm}
\overline{\not{\hspace{-.1cm} \partial}} \, \partial^2 \Bigr)
\Bigl[ \frac{\ln(\mu^2 \Delta x^2)}{\Delta x^2} \Bigr]} \nonumber \\
& & \hspace{.5cm} + \Bigl(-8 \; \hspace{-.1cm} \overline{\not{\hspace{-.1cm}
\partial}} \partial^2 \!+\! 4 \hspace{-.1cm} \not{\hspace{-.1cm} \partial}
\nabla^2 \Bigr) \Bigl[ \frac{\ln(\frac14 H^2\Delta x^2)}{\Delta x^2}
\Bigr] \!+\! 7 \hspace{-.1cm} \not{\hspace{-.1cm} \partial} \, \nabla^2 
\Bigl[ \frac1{\Delta x^2} \Bigr]\!\Biggr\} + O(\kappa^4) \; , \qquad
\end{eqnarray}
with the replacement,
\begin{equation}
\Delta x^2(x;x') \longrightarrow \Delta x^2_{\scriptscriptstyle +-}(x;x')
\equiv \Vert \vec{x} - \vec{x}' \Vert^2 - (\e-\e' + i\d)^2 \; . \label{D+-}
\end{equation}}
\end{itemize}
The difference of the ${\scriptscriptstyle ++}$ and ${\scriptscriptstyle +-}$
terms leads to zero contribution in (\ref{Diraceqn}) unless the point
$x^{\prime \mu}$ lies on or within the past light-cone of $x^{\mu}$.

We can only solve for the one loop corrections to the field because we lack
the higher loop contributions to the self-energy. The general perturbative
expansion takes the form,
\begin{equation}
\Psi(x) = \sum_{\ell = 0}^{\infty} \kappa^{2\ell} \Psi^{\ell}(x)
\qquad {\rm and}\,\,\, \Bigl[\Sigma\Bigr](x;x') =
\sum_{\ell=1}^{\infty} \kappa^{2\ell}
\Bigl[\Sigma^{\ell}\Bigr](x;x') \; .
\end{equation}
One substitutes these expansions into the effective Dirac equation
(\ref{Diraceqn}) and then segregates powers of $\kappa^2$,
\begin{equation}
i\hspace{-.1cm}\not{\hspace{-.08cm} \partial} \Psi^0(x) = 0 \qquad , \qquad
i\hspace{-.1cm}\not{\hspace{-.08cm} \partial} \Psi^1(x) = \int d^4x'
\Bigl[\Sigma^1\Bigr](x;x') \Psi^0(x') \qquad {\rm et\ cetera.}
\end{equation}
We shall work out the late time limit of the one loop correction
$\Psi^1_i(\eta,\vec{x};\vec{k},s)$ for a spatial plane wave of helicity
$s$,
\begin{equation}
\Psi^0_i(\eta,\vec{x};\vec{k},s) = \frac{e^{-i k \eta}}{\sqrt{2k}}
u_i(\vec{k},s) e^{i \vec{k} \cdot \vec{x}} \qquad {\rm where} \qquad k^{\ell}
\gamma^{\ell}_{ij} u_j(\vec{k},s) = k \gamma^0_{ij} u_j(\vec{k},s)
\; .\label{freefun}
\end{equation}

In the next section we derive the effective field equation. In
section 3 we derive some key simplifications. In section 4 we solve
for the late time limit of $\Psi^1_i(\eta,\vec{x};\vec{k},s)$. The
result takes the surprising form of a time dependent field strength
renormalization of the tree order solution. In section 5 we show
that this can be understood qualitatively using mean field theory.
Our results are summarized and discussed in section 6.

\section{The Effective Field Equations}

The purpose of this section is to elucidate the relation between the 
Heisenberg operators of Dirac + Einstein --- $\overline{\psi}_i(x)$,
$\psi_i(x)$ and $h_{\mu\nu}(x)$ --- and the $\comp$-number plane wave 
mode solutions $\Psi_i(x;\vec{k},s)$ of the linearized, effective 
Dirac equation (\ref{Diraceqn}). After explaining the relation we
work out an example, at one loop order, in a simple scalar analogue 
model. Finally, we return to Dirac + Einstein to explain how 
$\Psi_i(x;\vec{k},s)$ changes with variations of the gauge.

\subsection{Heisenberg operators and effective field equations}

The invariant Lagrangian of Dirac + Einstein in $D$ spacetime dimensions
is,
\begin{equation}
\mathcal{L} = \frac1{16 \pi G} \Bigl(R \!-\! (D\!-\!1) (D\!-\!2) H^2\Bigr) 
\sqrt{-g} \!+\! \overline{\psi} e^{\mu}_{~b} \gamma^b \Bigl(i\partial_{\mu} 
\!-\! \frac12 A_{\mu cd} J^{cd}\Bigr) \psi \sqrt{-g} \; .
\end{equation}
Here $e_{\mu b}$ is the vierbein field and $g_{\mu\nu} \equiv e_{\mu b} 
e_{\nu c} \eta^{bc}$ is the metric. The metric and vierbein-compatible 
connections are,
\begin{equation}
\Gamma^{\rho}_{~\mu\nu} \equiv \frac12 g^{\rho\sigma} \Bigl(g_{\sigma \mu ,
\nu} + g_{\nu \sigma , \mu} - g_{\mu \nu , \sigma}\Bigr) \qquad {\rm and}
\qquad A_{\mu cd} \equiv e^{\nu}_{~c} \Bigl( e_{\nu d , \mu} - 
\Gamma^{\rho}_{~\mu\nu} e_{\rho d}\Bigr) \; .
\end{equation}
The Ricci scalar is,
\begin{equation}
R \equiv g^{\mu\nu} \Bigl( \Gamma^{\rho}_{~\nu\mu , \rho} - \Gamma^{\rho}_{
~\rho \mu , \nu} + \Gamma^{\rho}_{~\rho \sigma} \Gamma^{\sigma}_{~\nu\mu}
- \Gamma^{\rho}_{~\nu \sigma} \Gamma^{\sigma}_{~\rho \mu} \Bigr) \; .
\end{equation}
The gamma matrices $\gamma^b_{ij}$ have spinor indices $i, j \in \{1,2,3,4\}$
and obey the usual anti-commutation relations,
\begin{equation}
\{\gamma^b , \gamma^c\} = -2 \eta^{bc} I \; .
\end{equation}
The Lorentz generators of the bispinor representation are,
\begin{equation}
J^{bc} \equiv \frac{i}4 [\gamma^b ,\gamma^c] \; .
\end{equation}

We employ the Lorentz symmetric gauge, $e_{\mu b} = e_{b \mu}$,
which permits one to perturbatively determine the vierbein in
terms of the metric and their respective backgrounds (denoted with
overlines) \cite{RPW2},
\begin{equation}
e_{\mu b}[g] = \Bigl(\sqrt{g {\overline{g}}_0^{-1}} \Bigr)_{\mu}^{~\nu}
\; \overline{e}_{\nu b} \; .
\end{equation}
We define the graviton field $h_{\mu\nu}$ in de Sitter conformal 
coordinates as follows,
\begin{equation}
g_{\mu\nu}(x) \equiv a^2 \Bigl( \eta_{\mu\nu} + \kappa h_{\mu\nu}(x)\Bigr)
\qquad {\rm where} \qquad a = -\frac1{H \eta} \; .
\end{equation}
By convention the indices of $h_{\mu\nu}$ are raised and lowered with
the Lorentz metric. We fix the general coordinate freedom by adding the
gauge fixing term,
\begin{equation}
\mathcal{L}_{\scriptscriptstyle {\rm GF}} = -\frac12 a^{D-2} \eta^{\mu\nu} 
F_{\mu} F_{\nu} \quad {\rm where} \quad F_{\mu} = \eta^{\rho\sigma} \Bigl( 
h_{\mu\rho , \sigma} \!-\! \frac12 h_{\rho \sigma , \mu} \!+\! (D\!-\!2) H a 
h_{\mu \rho} \delta^0_{\sigma}\Bigr) .
\end{equation}

One solves the gauge-fixed Heisenberg operator equations perturbatively,
\begin{eqnarray}
h_{\mu\nu}(x) & = & h^0_{\mu\nu}(x) + \kappa h^1_{\mu\nu}(x) + \kappa^2
h^2_{\mu\nu}(x) + \ldots \; , \qquad \\
\psi_i(x) & = & \psi^0_i(x) + \kappa \psi^1_i(x) + \kappa^2 \psi^2_i(x) + 
\ldots \; .
\end{eqnarray}
Because our state is released in free vacuum at $t=0$ ($\eta = -1/H$), 
it makes sense to express the operator as a functional of the creation
and annihilation operators of this free state. So our initial conditions
are that $h_{\mu\nu}$ and its first time derivative coincide with those of 
$h^0_{\mu\nu}(x)$ at $t=0$, and also that $\psi_i(x)$ coincides with 
$\psi^0_i(x)$. The zeroth order solutions to the Heisenberg field equations 
take the form,
\begin{eqnarray}
h^0_{\mu\nu}(x) & = & \int \frac{d^{D-1}k}{(2\pi)^{D-1}} \sum_{\lambda}
\Bigl\{ \epsilon_{\mu\nu}(\eta;\vec{k},\lambda) e^{i \vec{k} \cdot \vec{x}}
\alpha(\vec{k},\lambda) \nonumber \\
& & \hspace{5cm} + \epsilon^*_{\mu\nu}(\eta;\vec{k},\lambda) 
e^{-i \vec{k} \cdot \vec{x}} \alpha^{\dagger}(\vec{k},\lambda) \Bigr\} \; , 
\qquad \\
\psi^0_i(x) & = & a^{-(\frac{D-1}2)} \int \frac{d^{D-1}k}{(2\pi)^{D-1}} 
\sum_s \Bigl\{ \frac{e^{-ik \eta}}{\sqrt{2 k}} u_i(\vec{k},s) e^{i \vec{k} 
\cdot \vec{x}} b(\vec{k},s) \nonumber \\
& & \hspace{5cm} + \frac{e^{ik \eta}}{\sqrt{2 k}} v_i(\vec{k},\lambda) 
e^{-i \vec{k} \cdot \vec{x}} c^{\dagger}(\vec{k},s) \Bigr\} \; . \qquad
\end{eqnarray}
The graviton mode functions are proportional to Hankel functions whose 
precise specification we do not require. The Dirac mode functions $u_i(\vec{k},
s)$ and $v_i(\vec{k},s)$ are precisely those of flat space by
virtue of the conformal invariance of massless fermions. The canonically
normalized creation and annihilation operators obey,
\begin{eqnarray}
\Bigl[\alpha(\vec{k},\lambda), \alpha^{\dagger}(\vec{k}',\lambda')\Bigr]
& = & \delta_{\lambda \lambda'} (2\pi)^{D-1} \delta^{D-1}(\vec{k} \!-\! 
\vec{k}') \label{alpop} \; , \\
\Bigl\{b(\vec{k},s), b^{\dagger}(\vec{k}',s')\Bigr\} & = & \delta_{s s'} 
(2\pi)^{D-1} \delta^{D-1}(\vec{k} \!-\!  \vec{k}') =
\Bigl\{c(\vec{k},s), c^{\dagger}(\vec{k}',s')\Bigr\} \; . \qquad \label{bcop}
\end{eqnarray}

The zeroth order Fermi field $\psi^0_i(x)$ is an anti-commuting operator
whereas the mode function $\Psi^0(x;\vec{k},s)$ is a $\comp$-number. The latter 
can be obtained from the former by anti-commuting with the fermion creation
operator,
\begin{equation}
\Psi^0_i(x;\vec{k},s) = a^{\frac{D-1}2} \Bigl\{\psi^0_i(x), 
b^{\dagger}(\vec{k},s)\Bigr\} = \frac{e^{-i k \eta}}{\sqrt{2k}} u_i(\vec{k},s)
e^{i \vec{k} \cdot \vec{x}} \; .
\end{equation}
The higher order contributions to $\psi_i(x)$ are no longer linear in the
creation and annihilation operators, so anti-commuting the full solution
$\psi_i(x)$ with $b^{\dagger}(\vec{k},s)$ produces an operator. The 
quantum-corrected fermion mode function we obtain by solving (\ref{Diraceqn})
is the expectation value of this operator in the presence of the state which
is free vacuum at $t=0$,
\begin{equation}
\Psi_i(x;\vec{k},s) = a^{\frac{D-1}2} \Bigl\langle \Omega \Bigl\vert 
\Bigl\{ \psi_i(x), b^{\dagger}(\vec{k},s) \Bigr\} \Bigr\vert \Omega 
\Bigr\rangle \; . \label{SKop}
\end{equation}
This is what the Schwinger-Keldysh field equations give. The more familiar,
in-out effective field equations obey a similar relation except that one
defines the free fields to agree with the full ones in the asymptotic past,
and one takes the in-out matrix element after anti-commuting.

\subsection{A worked-out example}

It is perhaps worth seeing a worked-out example, at one loop order, of the 
relation (\ref{SKop}) between the Heisenberg operators and the 
Schwinger-Keldysh field equations. To simplify the analysis we will work 
with a model of two scalars in flat space,
\begin{equation}
\mathcal{L} = - \partial_{\mu} \varphi^* \partial^{\mu} \varphi -
m^2 \varphi^* \varphi - \lambda \chi \!:\! \varphi^* \varphi \!:\! -
\frac12 \partial_{\mu} \chi \partial^{\mu} \chi \; . \label{2sL}
\end{equation}
In this model $\varphi$ plays the role of our fermion $\psi_i$, and 
$\chi$ plays the role of the graviton $h_{\mu\nu}$. Note that we have
normal-ordered the interaction term to avoid the harmless but time-consuming
digression that would be required to deal with $\chi$ developing a nonzero
expectation value. We shall also omit discussion of counterterms.

The Heisenberg field equations for (\ref{2sL}) are,
\begin{eqnarray}
\partial^2 \chi - \lambda \!:\! \varphi^* \varphi \!:\! & = & 0 \; , \\
(\partial^2 - m^2) \varphi - \lambda \chi \varphi & = & 0 \; .
\end{eqnarray}
As with Dirac + Einstein, we solve these equations perturbatively,
\begin{eqnarray}
\chi(x) & = & \chi^0(x) + \lambda \chi^1(x) + \lambda^2 \chi^2(x) + 
\ldots \; , \\
\varphi(x) & = & \varphi^0(x) + \lambda \varphi^1(x) + \lambda^2 \varphi^2(x) 
+ \ldots \; .
\end{eqnarray}
The zeroth order solutions are,
\begin{eqnarray}
\chi^0(x) & = & \int \frac{d^{D-1}k}{(2\pi)^{D-1}} \Bigl\{ \frac{e^{-ikt}}{
\sqrt{2k}} e^{i \vec{k} \cdot \vec{x}} \alpha(\vec{k}) + \frac{e^{ikt}}{
\sqrt{2k}} e^{-i \vec{k} \cdot \vec{x}} \alpha^{\dagger}(\vec{k}) \Bigr\} 
\; , \\
\varphi^0(x) & = & \int \frac{d^{D-1}k}{(2\pi)^{D-1}} \Bigl\{ \frac{e^{-i
\omega t}}{\sqrt{2 \omega}} e^{i \vec{k} \cdot \vec{x}} b(\vec{k}) + 
\frac{e^{i \omega t}}{\sqrt{2 \omega}} e^{-i \vec{k} \cdot \vec{x}} 
c^{\dagger}(\vec{k}) \Bigr\} \; .
\end{eqnarray}
Here $k \equiv \Vert \vec{k} \Vert$ and $\omega \equiv \sqrt{k^2 + m^2}$.
The creation and annihilation operators are canonically normalized,
\begin{equation}
\Bigl[\alpha(\vec{k}),\alpha^{\dagger}(\vec{k}')\Bigr] = 
\Bigl[b(\vec{k}),b^{\dagger}(\vec{k}')\Bigr] = 
\Bigl[c(\vec{k}),c^{\dagger}(\vec{k}')\Bigr] = (2\pi)^{D-1} \delta^{D-1}(
\vec{k} - \vec{k}') \; .
\end{equation}
We choose to develop perturbation theory so that all the operators and their
first time derivatives agree with the zeroth order solutions at $t=0$. The
first few higher order terms are,
\begin{eqnarray}
\chi^1(x) & \!\!\!\!\! = \!\!\!\!\! & \int_0^t \!\! dt' \!\! \int d^{D-1}x' \,
\Bigl\langle x \Bigl\vert \frac1{\partial^2} \Bigr\vert x' 
\Bigr\rangle_{\rm ret} \!:\! \varphi^{0*}(x') \varphi^0(x') \!:\! \; , \\
\varphi^1(x) & \!\!\!\!\! = \!\!\!\!\! & \int_0^t \!\! dt' \!\! \int d^{D-1}x' 
\, \Bigl\langle x \Bigl\vert \frac1{\partial^2 \!-\! m^2} \Bigr\vert x' 
\Bigr\rangle_{\rm ret} \chi^0(x') \varphi^0(x') \; , \\
\varphi^2(x) & \!\!\!\!\! = \!\!\!\!\! & \int_0^t \!\! dt' \!\! \int d^{D-1}x' 
\Bigl\langle x \Bigl\vert \frac1{\partial^2 \!-\! m^2} \Bigr\vert x' 
\Bigr\rangle_{\rm ret} \Bigl\{\chi^1(x') \varphi^0(x') \!+\! \chi^0(x') 
\varphi^1(x') \Bigr\} . \qquad
\end{eqnarray}

The commutator of $\varphi^0(x)$ with $b^{\dagger}(\vec{k})$ is a
$\comp$-number,
\begin{equation}
\Bigl[ \varphi^0(x) , b^{\dagger}(\vec{k}) \Bigr] = \frac{e^{-i \omega t}}{
\sqrt{2 \omega}} \, e^{i \vec{k} \cdot \vec{x}} \equiv \Phi^0(x;\vec{k}) \; .
\label{Phi^0}
\end{equation}
However, commuting the full solution with $b^{\dagger}(\vec{k})$ leaves
operators,
\begin{eqnarray}
\lefteqn{\Bigl[ \varphi(x) , b^{\dagger}(\vec{k}) \Bigr] = \Phi^0(x;\vec{k})
+ \lambda \int_0^t \!\! dt' \!\! \int d^{D-1}x' \,
\Bigl\langle x \Bigl\vert \frac1{\partial^2 \!-\! m^2} \Bigr\vert x' 
\Bigr\rangle_{\rm ret} \chi^0(x') \Phi^0(x';\vec{k}) } \nonumber \\
& & \hspace{-.5cm} + \lambda^2 \int_0^t \!\! dt' \!\! \int d^{D-1}x' \,
\Bigl\langle x \Bigl\vert \frac1{\partial^2 \!-\! m^2} \Bigr\vert x' 
\Bigr\rangle_{\rm ret} \Biggl\{ \Bigl[\chi^1(x') , b^{\dagger}(\vec{k})\Bigr]
\varphi^0(x') + \chi^1(x') \Phi^0(x';\vec{k}) \nonumber \\
& & \hspace{6cm} + \chi^0(x') \Bigl[ \varphi^1(x') , b^{\dagger}(\vec{k}) 
\Bigr] \Biggr\} + O(\lambda^3) \; . \qquad \label{com} 
\end{eqnarray}
The commutators in (\ref{com}) are easily evaluated,
\begin{eqnarray}
\lefteqn{\Bigl[\chi^1(x') , b^{\dagger}(\vec{k})\Bigr] \varphi^0(x') }
\nonumber \\
& & \hspace{1.5cm} = \int_0^{t'} \!\! dt'' \!\! \int d^{D-1}x'' \, \Bigl\langle 
x' \Bigl\vert \frac1{\partial^2} \Bigr\vert x'' \Bigr\rangle_{\rm ret} 
\varphi^{0*}(x'') \varphi^0(x') \Phi^0(x'';\vec{k}) \; , \qquad \\
\lefteqn{\chi^0(x') \Bigl[\varphi^1(x') , b^{\dagger}(\vec{k}) \Bigr] }
\nonumber \\ 
& & \hspace{1.5cm} = \int_0^{t'} \!\! dt''\!\!\int d^{D-1}x'' \, \Bigl\langle x'
\Bigl\vert \frac1{\partial^2 \!-\! m^2} \Bigr\vert x'' \Bigr\rangle_{\rm ret} 
\chi^0(x') \chi^0(x'') \Phi^0(x'';\vec{k}) \; . \qquad
\end{eqnarray}
Hence the expectation value of (\ref{com}) gives,
\begin{eqnarray}
\lefteqn{\Bigl\langle \Omega \Bigl\vert \Bigl[ \varphi(x), b^{\dagger}(\vec{k}) 
\Bigr] \Bigr\vert \Omega \Bigr\rangle = \Phi^0(x;\vec{k}) + \lambda^2 \int_0^t 
\!\! dt' \!\! \int d^{D-1}x' \, \Bigl\langle x \Bigl\vert \frac1{\partial^2 
\!-\! m^2} \Bigr\vert x' \Bigr\rangle_{\rm ret} } \nonumber \\
& & \times \int_0^{t'} \!\! dt'' \!\! \int d^{D-1}x'' \, \Biggl\{ \Bigl\langle 
x' \Bigl\vert \frac1{\partial^2} \Bigr\vert x'' \Bigr\rangle_{\rm ret} 
\Bigl\langle \Omega \Bigl\vert \varphi^{0*}(x'') \varphi^0(x') \Bigr\vert
\Omega \Bigr\rangle \nonumber \\
& & \hspace{1.3 cm} + \Bigl\langle x' \Bigl\vert \frac1{\partial^2 \!-\! m^2} 
\Bigr\vert x'' \Bigr\rangle_{\rm ret} \Bigl\langle \Omega \Bigl\vert 
\chi^0(x') \chi^0(x'') \Bigr\vert \Omega \Bigr\rangle \Biggr\} \Phi^0(x'';
\vec{k}) + O(\lambda^4) \; . \qquad \label{expcom}
\end{eqnarray}

To make contact with the effective field equations we must first recognize
that the retarded Green's functions can be written in terms of expectation
values of the free fields,
\begin{eqnarray}
\lefteqn{\Bigl\langle x' \Bigl\vert \frac1{\partial^2} \Bigr\vert x'' 
\Bigr\rangle_{\rm ret} = -i \theta(t' \!-\! t'') \Bigl[ \chi^0(x') , 
\chi^0(x'')\Bigr] } \\
& & \hspace{1cm} = -i \theta(t' \!-\! t'') \Biggl\{ \Bigl\langle \Omega 
\Bigl\vert \chi^0(x') \chi^0(x'') \Bigr\vert \Omega \Bigr\rangle - \Bigl\langle
\Omega \Bigl\vert \chi^0(x'') \chi^0(x') \Bigr\vert \Omega \Bigr\rangle 
\Biggr\} \; , \qquad \\
\lefteqn{\Bigl\langle x' \Bigl\vert \frac1{\partial^2 \!-\! m^2} \Bigr\vert x'' 
\Bigr\rangle_{\rm ret} = -i \theta(t' \!-\! t'') \Bigl[ \varphi^0(x') , 
\varphi^{0*}(x'')\Bigr] } \\
& & \hspace{1cm} = -i \theta(t' \!-\! t'') \Biggl\{ \Bigl\langle \Omega 
\Bigl\vert \varphi^0(x') \varphi^{0*}(x'') \Bigr\vert \Omega \Bigr\rangle - 
\Bigl\langle \Omega \Bigl\vert \varphi^{*0}(x'') \varphi^0(x') \Bigr\vert 
\Omega \Bigr\rangle \Biggr\} \; . \qquad
\end{eqnarray}
Substituting these relations into (\ref{expcom}) and canceling some terms
gives the expression we have been seeking,
\begin{eqnarray}
\lefteqn{\Bigl\langle \Omega \Bigl\vert \Bigl[ \varphi(x), b^{\dagger}(\vec{k}) 
\Bigr] \Bigr\vert \Omega \Bigr\rangle = \Phi^0(x;\vec{k}) -i \lambda^2 \int_0^t 
\!\! dt' \!\! \int d^{D-1}x' \, \Bigl\langle x \Bigl\vert \frac1{\partial^2 
\!-\! m^2} \Bigr\vert x' \Bigr\rangle_{\rm ret} } \nonumber \\
& & \times \int_0^{t'} \!\! dt'' \!\! \int d^{D-1}x'' \, \Biggl\{ \Bigl\langle 
\Omega \Bigl\vert \chi^0(x') \chi^0(x'') \Bigr\vert \Omega \Bigr\rangle 
\Bigl\langle \Omega \Bigl\vert \varphi^0(x') \varphi^{0*}(x'') \Bigr\vert 
\Omega \Bigr\rangle \nonumber \\
& & \hspace{.9cm} - \Bigl\langle \Omega \Bigl\vert \chi^0(x'') \chi^0(x') 
\Bigr\vert \Omega \Bigr\rangle \Bigl\langle \Omega \Bigl\vert \varphi^{0*}(x'') 
\varphi^0(x') \Bigr\vert \Omega \Bigr\rangle \Biggr\} \Phi^0(x'';\vec{k}) 
+ O(\lambda^4) \; . \qquad \label{fexp}
\end{eqnarray}

We turn now to the effective field equations of the Schwinger-Keldysh 
formalism. The $\comp$-number field corresponding to $\varphi(x)$ at
linearized order is $\Phi(x)$. If the state is released at $t=0$ then the 
equation $\Phi(x)$ obeys is,
\begin{equation}
(\partial^2 - m^2) \Phi(x) - \int_0^t \!\! dt' \!\! \int d^{D-1}x' 
\Bigl\{ M^2_{\scriptscriptstyle ++}(x;x') + M^2_{\scriptscriptstyle +-}(x;x') 
\Bigr\} \Phi(x') = 0 \; . \label{Phieqn}
\end{equation}
The one loop diagram for the self-mass-squared of $\varphi$ is depicted in
Fig.~1.
\begin{center}
\begin{picture}(300,70)(0,0)
\DashCArc (150,20)(40,0,180){5} \ArrowLine(190,20)(110,20)
\ArrowLine(110,20)(50,20) \Vertex(110,20){3} \Text(110,10)[b]{$x$}
\ArrowLine(250,20)(190,20) \Vertex(190,20){3} \Text(191,10)[b]{$x'$}
\end{picture}
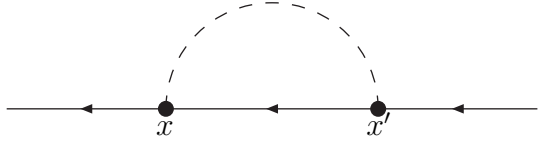
\\ {\rm Fig.~1: Self-mass-squared for $\varphi$ at one loop order. Solid lines 
stands for $\varphi$ propagators while dashed lines represent $\chi$ 
propagators.}
\end{center}
Because the self-mass-squared has two external lines, there are $2^2 = 4$
polarities in the Schwinger-Keldysh formalism. The two we require are
\cite{DW,FW},
\begin{eqnarray}
-i M^2_{\scriptscriptstyle ++}(x;x') & = & (-i \lambda)^2 \Bigl\langle x 
\Bigl\vert \frac{i}{\partial^2} \Bigr\vert x' \Bigr\rangle_{\scriptscriptstyle 
++} \Bigl\langle x \Bigl\vert \frac{i}{\partial^2 \!-\! m^2} \Bigr\vert x' 
\Bigr\rangle_{\scriptscriptstyle ++} + O(\lambda^4) \; , \label{M++} \\
-i M^2_{\scriptscriptstyle +-}(x;x') & = & (-i \lambda) (+i \lambda) 
\Bigl\langle x \Bigl\vert \frac{i}{\partial^2} \Bigr\vert x' \Bigr\rangle_{
\scriptscriptstyle +-} \Bigl\langle x \Bigl\vert \frac{i}{\partial^2 \!-\! m^2}
\Bigr\vert x' \Bigr\rangle_{\scriptscriptstyle +-} + O(\lambda^4) \; . \qquad
\label{M+-}
\end{eqnarray}

To recover (\ref{fexp}) we must express the various Schwinger-Keldysh
propagators in terms of expectation values of the free fields. The 
${\scriptscriptstyle ++}$ polarity gives the usual Feynman propagator
\cite{FW},
\begin{eqnarray}
\lefteqn{\Bigl\langle x \Bigl\vert \frac{i}{\partial^2} \Bigr\vert x' 
\Bigr\rangle_{\scriptscriptstyle ++} 
= \theta(t \!-\! t') \Bigl\langle \Omega \Bigl\vert
\chi^0(x) \chi^0(x') \Bigr\vert \Omega \Bigr\rangle \!+\! \theta(t' \!-\! t)
\Bigl\langle \Omega \Bigl\vert \chi^0(x') \chi^0(x) \Bigr\vert \Omega 
\Bigr\rangle \; , \qquad } \\
\lefteqn{\Bigl\langle x \Bigl\vert \frac{i}{\partial^2 \!-\! m^2} \Bigr\vert x' 
\Bigr\rangle_{\scriptscriptstyle ++} } \nonumber \\
& & \hspace{1.3cm} = \theta(t \!-\! t') \Bigl\langle \Omega 
\Bigl\vert \varphi^0(x) \varphi^{0*}(x') \Bigr\vert \Omega \Bigr\rangle \!+\! 
\theta(t' \!-\! t) \Bigl\langle \Omega \Bigl\vert \varphi^{0*}(x') \varphi^0(x) 
\Bigr\vert \Omega \Bigr\rangle \; . \qquad
\end{eqnarray}
The ${\scriptscriptstyle +-}$ polarity propagators are \cite{FW},
\begin{eqnarray}
\Bigl\langle x \Bigl\vert \frac{i}{\partial^2} \Bigr\vert x' 
\Bigr\rangle_{\scriptscriptstyle +-} 
& = & \Bigl\langle \Omega \Bigl\vert \chi^0(x') \chi^0(x) \Bigr\vert \Omega 
\Bigr\rangle \; , \\
\Bigl\langle x \Bigl\vert \frac{i}{\partial^2 \!-\! m^2} \Bigr\vert x' 
\Bigr\rangle_{\scriptscriptstyle +-} 
& = & \Bigl\langle \Omega \Bigl\vert \varphi^{0*}(x') \varphi^0(x) \Bigr\vert 
\Omega \Bigr\rangle \; . \qquad
\end{eqnarray}
Substituting these relations into (\ref{M++}-\ref{M+-}) and making use of the
identity $1 = \theta(t \!-\! t') \!+\! \theta(t' \!-\! t)$ gives,
\begin{eqnarray}
\lefteqn{ M^2_{\scriptscriptstyle ++}(x;x') + M^2_{\scriptscriptstyle +-}(x;x')
= -i \lambda^2 \theta(t \!-\! t') \Biggl\{
\Bigl\langle \Omega \Bigl\vert \chi^0(x) \chi^0(x') \Bigr\vert \Omega 
\Bigr\rangle } \nonumber \\
& & \hspace{-.7cm} \times \Bigl\langle \Omega \Bigl\vert \varphi^0(x) 
\varphi^{0*}(x') \Bigr\vert \Omega \Bigr\rangle \!-\! \Bigl\langle \Omega 
\Bigl\vert \chi^0(x') \chi^0(x) \Bigr\vert \Omega \Bigr\rangle \Bigl\langle 
\Omega \Bigl\vert \varphi^{0*}(x') \varphi^0(x) \Bigr\vert \Omega \Bigr\rangle 
\!\Biggr\} \!+\! O(\lambda^4) \, . \qquad \label{oneloop}
\end{eqnarray}

We now solve (\ref{Phieqn}) perturbatively. The free plane wave mode function 
(\ref{Phi^0}) is of course a solution at order $\lambda^0$. With 
(\ref{oneloop}) we easily recognize its perturbative development as,
\begin{eqnarray}
\lefteqn{\Phi(x;\vec{k}) = \Phi^0(x;\vec{k}) -i \lambda^2 \int_0^t 
\!\! dt' \!\! \int d^{D-1}x' \, \Bigl\langle x \Bigl\vert \frac1{\partial^2 
\!-\! m^2} \Bigr\vert x' \Bigr\rangle_{\rm ret} } \nonumber \\
& & \times \int_0^{t'} \!\! dt'' \!\! \int d^{D-1}x'' \, \Biggl\{ \Bigl\langle 
\Omega \Bigl\vert \chi^0(x') \chi^0(x'') \Bigr\vert \Omega \Bigr\rangle 
\Bigl\langle \Omega \Bigl\vert \varphi^0(x') \varphi^{0*}(x'') \Bigr\vert 
\Omega \Bigr\rangle \nonumber \\
& & \hspace{.9cm} - \Bigl\langle \Omega \Bigl\vert \chi^0(x'') \chi^0(x') 
\Bigr\vert \Omega \Bigr\rangle \Bigl\langle \Omega \Bigl\vert \varphi^{0*}(x'')
\varphi^0(x') \Bigr\vert \Omega \Bigr\rangle \Biggr\} \Phi^0(x'';\vec{k}) 
+ O(\lambda^4) \; . \qquad 
\end{eqnarray}
That agrees with (\ref{fexp}), so we have established the desired connection,
\begin{equation}
\Phi(x;\vec{k}) = \Bigl\langle \Omega \Bigl\vert \Bigl[ \varphi(x), 
b^{\dagger}(\vec{k}) \Bigr] \Bigr\vert \Omega \Bigr\rangle \; ,
\end{equation}
at one loop order.

\subsection{The gauge issue}

The preceding discussion has made clear that we are working in a particular
local Lorentz and general coordinate gauge. We are also doing perturbation
theory. The function $\Psi^0_i(x;\vec{k},s)$ describes how a free fermion of
wave number $\vec{k}$ and helicity $s$ propagates through classical de Sitter 
background in our gauge. What $\Psi^1_i(x;\vec{k},s)$ gives is the first 
quantum correction to this mode function. It is natural to wonder how the
effective field $\Psi_i(x;\vec{k},s)$ changes if a different gauge is used.

The operators of the original, invariant Lagrangian transform as follows under 
diffeomorphisms ($x^{\mu} \rightarrow x^{\prime \mu}$) and local Lorentz 
rotations ($\Lambda_{ij}$),\footnote{Of course the spinor and vector
representations of the local Lorentz transformation are related as usual,
with same parameters $\omega_{cd}(x)$ contracted into the appropriate
representation matrices,
\begin{eqnarray}
\Lambda_{ij} \equiv \delta_{ij} - \frac{i}2 \omega_{cd} J^{cd}_{~~ij} + \ldots 
\qquad {\rm and} \qquad
\Lambda_b^{~c} = \delta_b^{~c} - \omega_b^{~c} + \ldots \; . \nonumber
\end{eqnarray}}
\begin{eqnarray}
\psi'_i(x) & = & \Lambda_{ij}\Bigl(x^{\prime -1}(x)\Bigr) \psi_j\Bigl(
x^{\prime -1}(x)\Bigr) \; , \\
e'_{\mu b}(x) & = & \frac{\partial x^{\nu}}{\partial x^{\prime \mu}}
\Lambda_b^{~c}\Bigl(x^{\prime -1}(x)\Bigr) e_{\nu c}\Bigl( x^{\prime -1}(x)
\Bigr) \; .
\end{eqnarray}
The invariance of the theory guarantees that the transformation of any
solution is also a solution. Hence the possibility of performing local 
transformations precludes the existence of a unique initial value solution.
This is why no Hamiltonian formalism is possible until the gauge has been
fixed sufficiently to eliminate transformations which leave the initial
value surface unaffected.

Different gauges can be reached using field-dependent gauge transformations 
\cite{TW7}. This has a relatively simple effect upon the Heisenberg operator 
$\psi_i(x)$, but a complicated one on the linearized effective field 
$\Psi_i(x;\vec{k},s)$. Because local Lorentz and diffeomorphism gauge 
conditions are typically specified in terms of the gravitational fields, we 
assume $x^{\prime \mu}$ and $\Lambda_{ij}$ depend upon the graviton field 
$h_{\mu\nu}$. Hence so too does the transformed field,
\begin{equation}
\psi'_i[h](x) = \Lambda_{ij}[h]\Bigl(x^{\prime -1}[h](x)\Bigr) 
\psi_j\Bigl(x^{\prime -1}[h](x)\Bigr) \; .
\end{equation}
In the general case that the gauge changes even on the initial value
surface, the creation and annihilation operators also transform,
\begin{equation}
b'[h](\vec{k},s) = \frac1{\sqrt{2k}} u^*_i(\vec{k},s) \int d^{D-1}x \,
e^{-i \vec{k} \cdot \vec{x}} \psi'_i[h](\eta_i,\vec{x}) \; ,
\end{equation}
where $\eta_i \equiv -1/H$ is the initial conformal time. Hence the
linearized effective field transforms to,
\begin{equation}
\Psi'_i(x;\vec{k},s) = a^{\frac{D-1}2} \Bigl\langle \Omega \Bigl\vert 
\Bigl\{ \psi'_i[h](x), b^{\prime\dagger}[h](\vec{k},s) \Bigr\} \Bigr\vert 
\Omega \Bigr\rangle \; . \label{SKprime}
\end{equation}
This is quite a complicated relation. Note in particular that the
$h_{\mu\nu}$ dependence of $x^{\prime \mu}[h]$ and $\Lambda_{ij}[h]$ means
that $\Psi'_i(x;\vec{k},s)$ is not simply a Lorentz transformation of the
original function $\Psi_i(x;\vec{k},s)$ evaluated at some transformed point.

\section{Some Key Reductions}

The purpose of this section is to derive three results that are used
repeatedly in reducing the nonlocal contributions to the effective field
equations.  We observe that the nonlocal terms of (\ref{ren}) contain
$1/\Delta x^2$. We can avoid denominators by extracting another derivative,
\begin{equation}
\f{1}{\D x^2}=\frac{\del^2}4 \ln(\D x^2) \qquad {\rm and} \qquad
\f{\ln(\D x^2)}{\D x^2} = \frac{\del^2}8 \Bigl[\ln^2(\D x^{2}) - 2
\ln(\D x^2) \Bigr] \; . \label{id2}
\end{equation}
The Schwinger-Keldysh field equations involve the difference of
${\scriptscriptstyle ++}$ and ${\scriptscriptstyle +-}$ terms, for example,
\bea
\hspace{-1cm} && \f{\ln(\mu^2\D x^2_{\scriptscriptstyle ++})}{\D x^2_{
\scriptscriptstyle ++}} - \f{\ln(\mu^2 \D x^2_{\scriptscriptstyle +-})}{\D
x^2_{\scriptscriptstyle +-}} \nn \\
\hspace{-1cm} & & = \f{\del^2}{8} \Biggl\{ \ln^2(\mu^2\D x^2_{\scriptscriptstyle
++}) - 2 \ln(\mu^2\D x^2_{\scriptscriptstyle ++}) - \ln^2(\mu^2\D x^2_{
\scriptscriptstyle +-})+2\ln(\mu^2\D x^2_{\scriptscriptstyle +-}) \Biggr\}.
\label{r1}
\eea
We now define the coordinate intervals $\D\eta \equiv \e \!-\! \e'$ and
$\D x\equiv \Vert \vec{x} \!-\! \vec{x}' \Vert$ in terms of which the
${\scriptscriptstyle ++}$ and ${\scriptscriptstyle +-}$ intervals are,
\be
\D x^2_{\scriptscriptstyle ++} = \D x^2 - (\vert \D\e \vert \!-\! i\d)^2
\,\,\, {\rm and} \,\,\, \D x^2_{\scriptscriptstyle +-} = \D x^2 - (\D\e \!+\!
i\d)^2 \,\, .
\ee
When $\e'>\e$ we have $\D x^2_{\scriptscriptstyle ++} = \D x^2_{
\scriptscriptstyle +-}$ , so the ${\scriptscriptstyle ++}$ and
${\scriptscriptstyle +-}$ terms in (\ref{r1}) cancel. This means there
is no contribution from the future. When $\e'<\e$ and $\D x>\D\e$ (past
spacelike separation) we can take $\delta = 0$,
\begin{equation}
\ln(\mu^2\D x^2_{\scriptscriptstyle ++}) = \ln[\mu^2(\D x^2 \!-\! \D \e^2 )]
= \ln(\mu^2\D x^2_{\scriptscriptstyle +-}) \qquad (\Delta x > \Delta \eta > 0)
\,\, . \label{ln1}
\end{equation}
So the ${\scriptscriptstyle ++}$ and ${\scriptscriptstyle +-}$ terms again
cancel. Only for $\eta'<\eta$ and $\D x<\D\eta$ (past timelike separation) are
the two logarithms different,
\begin{equation}
\ln(\mu^2\D x^2_{\scriptscriptstyle +\pm}) = \ln[\mu^2(\D\eta^2 \!-\! \D x^2)]
\pm i \pi \qquad (\Delta \eta > \Delta x > 0) \,\, .
\end{equation}
Hence equation (\ref{r1}) can be written as,
\bea
\f{\ln(\mu^2\D x^2_{\scriptscriptstyle ++})}{\D x^2_{\scriptscriptstyle ++}}
- \f{\ln(\mu^2 \D x^2_{\scriptscriptstyle +-})}{\D x^2_{\scriptscriptstyle +-}}
= \f{i \pi}2 \del^2 \Biggl\{\theta(\D\eta \!-\! \D x) \Bigl[\ln(\mu^2(\D\e^2
\!-\! \D x^2) \!-\! 1\Bigr] \Biggr\} . \label{r2}
\eea
This step shows that the Schwinger-Kledysh formalism is causal.

To integrate (\ref{r2}) up against the plane wave mode function (\ref{freefun})
we first pull the $x^{\mu}$ derivatives outside the integration, then make the
change of variables $\vec{x}' \!=\! \vec{x} \!+\! \vec{r}$ and perform the
angular integrals,
\begin{eqnarray}
\lefteqn{\int d^4x' \Biggl\{ \f{\ln(\mu^2\D x^2_{++})}{\D x^2_{++}} -
\f{\ln(\mu^2\D x^2_{+-})}{\D x^2_{+-}} \Biggr\} \Psi^0_i(\e',\vec{x},\vec{k},s)}
\nn \\
&& = \f{i 2\pi^2}{k} u_i(\vec{k},s) \del^2 e^{i\vec{k}\cdot\vec{x}}
\int^{\e}_{\e_{i}} d\e' \frac{e^{-ik \eta'}}{\sqrt{2k}} \!\! \int^{\D\e}_{0}
\!\!\! dr r \sin(kr) \Bigl\{ \ln[\mu^2(\D\e^2 \!-\! r^2)] \!-\! 1\Bigr\} \nn \\
&& = \f{i2\pi^2}{k \sqrt{2k}} e^{i\vec{k} \cdot \vec{x}} u_i(\vec{k},s) 
[-\del^2_{0} \!-\! k^2] \int^{\e}_{\e_{i}} \!\! d\e' e^{-ik \eta'} {\Delta
\eta}^2 \nn \\
&& \hspace{4.5cm} \times \int_{0}^{1} \!\!\! dz z \sin(\a z) \Bigl\{
\ln(1 \!-\! z^2) \!+\! 2 \ln(\f{\mu\a}{k}) \!-\! 1 \Bigr\} . \qquad \label{r3}
\end{eqnarray}
Here $\alpha \equiv k \Delta \eta$ and $\e_{i} \equiv -1/H$ is the initial
conformal time, corresponding to physical time $t=0$. The integral over $z$
is facilitated by the special function,
\begin{eqnarray}
\lefteqn{\xi(\a) \equiv \int_{0}^{1} \!\! dz z \sin(\a z) \ln(1 \!-\! z^2) =
\f{2}{\a^2} \sin(\a) - \f{1}{\a^2} \Bigl[\cos(\a) \!+\! \a \sin(\a) \Bigr] }
\nn \\
& & \hspace{2cm} \times \Bigl[{\rm si}(2\a) \!+\! \f{\pi}{2} \Bigr] +
\Bigl[\sin(\a) \!-\! \a \cos(\a)\Bigr] \Bigl[{\rm ci}(2\a) \!-\! \gamma \!-\!
\ln(\f{\a}{2})\Bigr] \,\, . \qquad
\end{eqnarray}
Here $\gamma$ is the Euler-Mascheroni constant and the sine and cosine
integrals are,
\begin{eqnarray}
{\rm si}(x) & \equiv & -\int_{x}^{\infty} \!\! dt \, \f{\sin(t)}{t} =
- \f{\pi}{2} + \int_{0}^{x} \!\! dt \, \f{\sin t}{t} \,\, , \\
{\rm ci}(x) & \equiv & -\int_{x}^{\infty} \!\! dt \, \f{\cos t}{t} =
\g + \ln(x) + \int_{0}^{x} \!\! dt \, \Bigl[\f{\cos(t) - 1}{t}\Bigr] \,\, .
\end{eqnarray}
After substituting the $\xi$ function and performing the elementary integrals,
(\ref{r3}) becomes,
\begin{eqnarray}
\lefteqn{\int d^4x' \Biggl\{ \f{\ln(\mu^2\D x^2_{++})}{\D x^2_{++}} -
\f{\ln(\mu^2\D x^2_{+-})}{\D x^2_{+-}} \Biggr\} \Psi^0_i(\e',\vec{x},\vec{k},s)
= \f{i 2 \pi^2}{k \sqrt{2k}} e^{i \vec{k} \cdot \vec{x}} u_i(\vec{k},s) } \nn \\
& & \hspace{.5cm} \times (\del^2_{k \eta} \!+\! 1) \hspace{-.1cm} \int_{\e_i}^{
\e} \hspace{-.3cm} d\e' e^{-ik\e'} \Biggl\{\a^2\xi(\a) \!+\! \Bigl[2\ln(\f{
\mu\a}{k}) \!-\! 1\Bigr]\Bigl[\sin(\a) \!-\! \a \cos(\a) \Bigr] \Biggr\} .
\qquad \label{r4}
\end{eqnarray}

One can see that the integrand is of order $\a^3 \ln(\alpha)$ for small
$\alpha$, which means we can pass the derivatives through the integral.
After some rearrangements, the first key identity emerges,
\begin{eqnarray}
\lefteqn{ \int d^4x' \Biggl\{ \frac{\ln(\mu^2 \Delta
x^2_{\scriptscriptstyle ++})}{\Delta x^2_{\scriptscriptstyle ++}} -
\frac{\ln(\mu^2 \Delta x^2_{ \scriptscriptstyle +-})}{\Delta
x^2_{\scriptscriptstyle +-}} \Biggr\}
\Psi^0(\eta',\vec{x}';\vec{k},s) } \nonumber \\
& & = -i 4 \pi^2 k^{-1} \Psi^0(\eta,\vec{x};\vec{k},s)
\int_{\eta_i}^{\eta}\!\!\! d\eta' e^{i k \Delta \eta}
\Biggl\{-\cos(k \Delta \eta) \int_{\scriptscriptstyle 0}^{2 k \Delta
\eta} \!\!\!\!\!\! dt \, \frac{\sin(t)}{t} \nonumber \\
& & \hspace{3cm} + \sin(k \Delta \eta) \Biggl[
\int_{\scriptscriptstyle 0}^{2 k \Delta \eta} \!\!\!\!\!\! dt \,
\Bigl(\frac{\cos(t) \!-\! 1}{t}\Bigr) \!+\! 2 \ln(2 \mu \Delta
\eta)\Biggr] \Biggr\} . \label{key1}
\end{eqnarray}
Note that we have written $e^{-i k \eta'} = e^{-i k \eta} \times e^{-i k
\Delta \eta}$ and extracted the first phase to reconstruct the full
tree order solution $\Psi^0(\eta,\vec{x};\vec{k},s) = \frac{e^{-i k \eta}}{
\sqrt{2k}} u_i(\vec{k},s) e^{i \vec{k} \cdot \vec{x}}$.

The second identity derives from acting a d'Alembertian on (\ref{key1}).
The d'Alembertian passes through the tree order solution to give,
\begin{equation}
\partial^2 \Psi^0(\eta,\vec{x};\vec{k},s) = \Psi^0(\eta,\vec{x};\vec{k},s)
\partial_{\eta} (\partial_{\eta} \!-\! 2 i k) \; .
\end{equation}
Because the integrand goes like $\alpha \ln(\alpha)$ for small $\alpha$, we
can pass the first derivative through the integral to give,
\begin{eqnarray}
\lefteqn{ \partial^2 \int d^4x' \Biggl\{ \frac{\ln(\mu^2 \Delta
x^2_{\scriptscriptstyle ++})}{\Delta x^2_{\scriptscriptstyle ++}} -
\frac{\ln(\mu^2 \Delta x^2_{ \scriptscriptstyle +-})}{\Delta
x^2_{\scriptscriptstyle +-}} \Biggr\}
\Psi^0(\eta',\vec{x}';\vec{k},s) } \nonumber \\
& & \hspace{1.5cm} = i 4 \pi^2 \Psi^0(\eta,\vec{x};\vec{k},s)
\partial_{\eta} \int_{\e_i}^{\e} \!\! d\e' \Biggl\{
\int_{\scriptscriptstyle 0}^{2\a} \!\! dt \Bigl(\f{e^{it} \!-\!
1}{t} \Bigr) + 2 \ln(\f{2\mu\a}{k}) \Biggr\} \,\, . \qquad
\end{eqnarray}
We can pass the final derivative through the first integral but, for the
second, we must carry out the integration. The result is our second key
identity,
\begin{eqnarray}
\lefteqn{ \partial^2 \int d^4x' \Biggl\{ \frac{\ln(\mu^2 \Delta
x^2_{ \scriptscriptstyle ++})}{\Delta x^2_{\scriptscriptstyle ++}} -
\frac{\ln(\mu^2 \Delta x^2_{ \scriptscriptstyle +-})}{\Delta
x^2_{\scriptscriptstyle +-}}
\Biggr\} \Psi^0(\eta',\vec{x}';\vec{k},s) } \nonumber \\
& & \hspace{1.5cm} = i 4 \pi^2 \Psi^0(\eta,\vec{x};\vec{k},s) \Biggl\{ 2
\ln\Bigl[\frac{2\mu}{H} (1 \!+\! H \eta)\Bigr] \!+\!
\int_{\eta_i}^{\eta} \!\!\! d\eta' \, \Bigl(\frac{e^{i 2 k \Delta
\eta} \!-\! 1}{\Delta \eta}\Bigr) \Biggr\} . \qquad \label{key2}
\end{eqnarray}

The final key identity is derived through the same procedures. Because
they should be familiar by now we simply give the result,
\begin{eqnarray}
\lefteqn{ \int d^4x' \Biggl\{ \frac1{\Delta x^2_{\scriptscriptstyle
++}} - \frac1{\Delta x^2_{\scriptscriptstyle +-}} \Biggr\}
\Psi^0(\eta',\vec{x}'; \vec{k},s) } \nonumber \\
& & \hspace{3cm} = - i 4 \pi^2 k^{-1} \Psi^0(\eta,\vec{x};\vec{k},s)
\int_{\e_i}^{\e} \!\! d\e' \, e^{ik\D\e}\sin(k\D\e) \,\, . \qquad \label{key3}
\end{eqnarray}

\section{Solving the Effective Dirac Equation}

In this section we first evaluate the various nonlocal contributions using
the three identities of the previous section. Then we evaluate the vastly
simpler and, as it turns out, more important, local contributions. Finally,
we solve for $\Psi^1(\eta,\vec{x};\vec{k},s)$ at late times.

The various nonlocal contributions to (\ref{Diraceqn}) take the form,
\begin{eqnarray}
\lefteqn{\int \!\! d^4x' \sum_{I=1}^5 U^I_{ij} \Biggr\{\f{\ln(\a_{I}^2\D x^2_{
\scriptscriptstyle ++})}{\D x^2_{\scriptscriptstyle ++}} - \f{\ln(\a_{I}^2\D
x^2_{\scriptscriptstyle +-})}{\D x^2_{\scriptscriptstyle +-}} \Biggl\}
\Psi^0_j(\eta',\vec{x}';\vec{k},s) } \nonumber \\
& & \hspace{4cm} + \int \!\! d^4x' U^6_{ij} \Biggr\{\f1{\D x^2_{
\scriptscriptstyle ++}} - \f1{\D x^2_{\scriptscriptstyle +-}} \Biggl\}
\Psi^0_j(\eta',\vec{x}';\vec{k},s) \,\, . \qquad \label{analyticform}
\end{eqnarray}
The spinor differential operators $U^I_{ij}$ are listed in
Table \ref{nond}. The constants $\a_{I}$ are $\mu$ for $I = 1,2,3$,
and $\f12 H$ for $I=4,5$.
\begin{table}

\vbox{\tabskip=0pt \offinterlineskip
\def\tablerule{\noalign{\hrule}}
\halign to390pt {\strut#& \vrule#\tabskip=1em plus2em& \hfil#&
\vrule#& \hfil#\hfil& \vrule#& \hfil#& \vrule#& \hfil#\hfil&
\vrule#\tabskip=0pt\cr
\tablerule
\omit&height4pt&\omit&&\omit&&\omit&&\omit&\cr &&\omit\hidewidth $I$
&&\omit\hidewidth $U^I_{ij}$ \hidewidth&&
\omit\hidewidth $I$ \hidewidth&& \omit\hidewidth $U^I_{ij}$ \hidewidth&\cr
\omit&height4pt&\omit&&\omit&&\omit&&\omit&\cr
\tablerule
\omit&height2pt&\omit&&\omit&&\omit&&\omit&\cr && 1 && $(H^2
aa')^{-1}\not{\hspace{-.08cm}\del}\del^4$ && 4 &&
$\hspace{0.5cm}-8\not{\hspace{-.08cm}\bar{\del}}\del^2\hspace{0.5cm}
$ &\cr \omit&height2pt&\omit&&\omit&&\omit&&\omit&\cr
\tablerule
\omit&height2pt&\omit&&\omit&&\omit&&\omit&\cr && 2 && $\f{15}{2}
\not{\hspace{-.08cm}\del}\del^2$ && 5 &&
$4\not{\hspace{-.08cm}\del}\nabla^2$ &\cr
\omit&height2pt&\omit&&\omit&&\omit&&\omit&\cr
\tablerule
\omit&height2pt&\omit&&\omit&&\omit&&\omit&\cr && 3 &&
$-\not{\hspace{-.08cm}\bar{\del}}\del^2$ && 6 &&
$7\not{\hspace{-.08cm}\del}\nabla^2$ &\cr
\omit&height2pt&\omit&&\omit&&\omit&&\omit&\cr \tablerule}}

\caption{Derivative operators $U^I_{ij}$: Their common prefactor is
$\f{\kappa^2 H^2}{2^8\pi^4}$.}

\label{nond}

\end{table}

As an example, consider the contribution from $U^2_{ij}$:
\begin{eqnarray}
\lefteqn{\f{15}{2} \f{\kappa^2H^2}{2^8\pi^4} \not{\hspace{-.08cm}\del} \del^2
\!\! \int \!\! d^4x' \Biggr\{\f{\ln(\mu^2\D x^2_{\scriptscriptstyle ++})}{\D
x^2_{\scriptscriptstyle ++}} - \f{\ln(\mu^2\D x^2_{\scriptscriptstyle +-})}{
\D x^2_{\scriptscriptstyle +-}}\Biggl\} \Psi^0(\eta',\vec{x}';\vec{k},s) }\nn\\
&& \hspace{-.5cm} = \f{15}{2} \f{\kappa^2H^2}{2^8\pi^4} \!\not{\hspace{-.1cm}
\del} \!\times\! i 4 \pi^2 \Psi^{0}(\e,\vec{x};\vec{k},s)\Biggl\{\!2\ln\Bigl[
\f{2\mu}{H}(1 \!+\! H\e) \Bigr] \!+\!\! \int_{\e_i}^{\e} \!\! d\e' \Bigl(\f
{e^{2ik\D\e} \!-\! 1}{\D\e} \Bigr)\! \Biggr\} , \qquad \\
&& \hspace{-.5cm} = \f{\kappa^2H^2}{2^6\pi^2} i H \g^0 \Psi^0(\eta,\vec{x};
\vec{k},s) \times \f{15}{2} \f{1}{
1 \!+\! H\e} \Biggl\{e^{2i\f{k}{H}(1+H\e)} \!+\! 1\Biggr\} . \label{non2}
\end{eqnarray}
In these reductions we have used 
$i \hspace{-.1cm} \not{\hspace{-.1cm}\del} \Psi^0(\eta,\vec{x};\vec{k},s) \!=\!
i \gamma^0 \Psi^0(\eta,\vec{x};\vec{k},s) \, \partial_{\eta}$ and (\ref{key2}).
Recall from the
Introduction that reliable predictions are only possible for late times, which
corresponds to $\eta \rightarrow 0^-$. We therefore take this limit,
\begin{eqnarray}
\lefteqn{\f{15}{2} \f{\kappa^2H^2}{2^8\pi^4} \not{\hspace{-.08cm}\del} \del^2
\!\! \int \!\! d^4x' \Biggr\{\f{\ln(\mu^2\D x^2_{\scriptscriptstyle ++})}{\D
x^2_{\scriptscriptstyle ++}} - \f{\ln(\mu^2\D x^2_{\scriptscriptstyle +-})}{
\D x^2_{\scriptscriptstyle +-}}\Biggl\} \Psi^0(\eta',\vec{x}';\vec{k},s)}\nn \\
& & \hspace{3cm} \longrightarrow \f{\kappa^2H^2}{2^6\pi^2} i H \g^0
\Psi^0(\eta,\vec{x};\vec{k},s) \times \f{15}{2} \Bigl\{\exp(2i\f{k}{H}) + 1
\Bigr\} . \qquad
\end{eqnarray}

The other five nonlocal terms have very similar reductions. Each of them also
goes to $\frac{\kappa^2 H^2}{2^6 \pi^2} \times i H \gamma^0 \Psi^0(\eta,
\vec{x};\vec{k},s)$ times a finite constant at late times. We summarize the
results in Table \ref{noncon} and relegate the details to an appendix.
\begin{table}

\vbox{\tabskip=0pt \offinterlineskip
\def\tablerule{\noalign{\hrule}}
\halign to390pt {\strut#& \vrule#\tabskip=1em plus2em& \hfil#\hfil&
\vrule#& \hfil#\hfil& \vrule#\tabskip=0pt\cr \tablerule
\omit&height6pt&\omit&&\omit&\cr && ${\rm I}$
&& Coefficient\ of\ the\ late\ time\ contribution\ from\ each\ $U^I_{ij}$ &\cr
\omit&height6pt&\omit&&\omit&\cr \tablerule
\omit&height4pt&\omit&&\omit&\cr && 1 &&
$0$ & \cr
\omit&height4pt&\omit&&\omit&\cr \tablerule
\omit&height4pt&\omit&&\omit&\cr && 2 &&
$\f{15}{2}\Bigl\{\exp(2i\f{k}{H})+1\Bigr\}$ & \cr
\omit&height4pt&\omit&&\omit&\cr \tablerule
\omit&height4pt&\omit&&\omit&\cr && 3 &&
$-i\f{k}{H}\Bigl\{2\ln(\f{2\mu}{H})-\int_{\e_i}^{0}d\e'
\Bigl(\f{\exp(-2ik\e')-1}{\e'}\Bigr)\Bigr\}$ & \cr
\omit&height4pt&\omit&&\omit&\cr \tablerule
\omit&height4pt&\omit&&\omit&\cr && 4 &&
$8i\f{k}{H}\int_{\e_i}^{0}d\e' \Bigl(\f{\exp(-2ik\e')-1}{\e'}\Bigr)$
& \cr \omit&height4pt&\omit&&\omit&\cr \tablerule
\omit&height4pt&\omit&&\omit&\cr && 5 &&
$4\f{k^2}{H}\int_{\e_i}^{0}d\e'e^{-2ik\e'}
\Bigl\{\int_{0}^{-2k\e'}dt\Bigl(\f{\exp(-it)-1}{t}\Bigr)
+2\ln(H\e')\Bigr\}$ & \cr \omit&height4pt&\omit&&\omit&\cr
\tablerule \omit&height4pt&\omit&&\omit&\cr && 6 &&
$-\f{7}{2}i\f{k}{H}\Bigl\{\exp(2i\f{k}{H})-1\Bigr\}$ & \cr
\omit&height4pt&\omit&&\omit&\cr \tablerule}}

\caption{Nonlocal contributions to $\int d^4x' [\Sigma](x;x') \Psi^0(\eta',
\vec{x'};\vec{k},s)$ at late times. Multiply each term by $\f{\kappa^2H^2}{2^6
\pi^2} \times iH \g^0 \Psi^0(\eta,\vec{x};\vec{k},s)$.}

\label{noncon}

\end{table}

The next step is to evaluate the local contributions. This is a straightforward
exercise in calculus, using only the properties of the tree order solution
(\ref{freefun}) and the fact that $\partial_{\mu} a = H a^2 \delta^0_{\mu}$.
The result is,
\begin{eqnarray}
\lefteqn{\f{i\kappa^2 H^2}{2^6\pi^2} \!\!\int \!\! d^4x' \Biggl\{ \f{\ln(aa')}{
H^2 aa'} \! \not{\hspace{-.1cm}\del} \del^2 \!+\! \f{15}{2} \ln(aa') \! \not{
\hspace{-.1cm}\del} \!-\! 7 \ln(aa') \!\! \not{\hspace{-.08cm} \bar{\del}}
\Biggr\} \d^4(x \!-\! x') \Psi^0(\eta',\vec{x}';\vec{k},s) } \nonumber \\
& & = \frac{i \kappa^2 H^2}{2^6 \pi^2} \Biggl\{ \frac{\ln(a)}{H^2 a}
\!\not{\hspace{-.1cm}\del} \del^2 \Bigl(\frac1{a} \Psi^0(\eta,\vec{x};\vec{k},s)
\Bigr) + \frac1{H^2 a} \! \not{\hspace{-.1cm}\del} \del^2 \Bigl(\frac{\ln(a)}{
a} \Psi^0(\eta,\vec{x};\vec{k},s)\Bigr) \nonumber \\
& & \hspace{1.1cm} + \frac{15}2 \Bigl(\ln(a) \! \not{\hspace{-.1cm}\del} \!+\!
\not{\hspace{-.1cm}\del} \ln(a)\Bigr) \Psi^0(\eta,\vec{x};\vec{k},s)
- 14 \ln(a) \!\not{\hspace{-.1cm}  \bar{\del}}
\Psi^0(\eta,\vec{x};\vec{k},s) \Biggr\} , \qquad \label{localgen} \\
& & = \frac{\kappa^2 H^2}{2^6 \pi^2} iH \gamma^0 \Psi^0(\eta,\vec{x};\vec{k},s)
\times \Biggl\{ \frac{17}2 a - 14 i \frac{k}{H} \ln(a) - 2 i \frac{k}{H}
\Biggr\} . \label{localtotal}
\end{eqnarray}

The local quantum corrections (\ref{localtotal}) are evidently much stronger
than their nonlocal counterparts in Table \ref{noncon}! Whereas the nonlocal
terms approach a constant, the leading local contribution grows like the
inflationary scale factor, $a = e^{H t}$. Even factors of $\ln(a)$ are
negligible by comparison. We can therefore write the late time limit
of the one loop field equation as,
\begin{eqnarray}
i\hspace{-.1cm}\not{\hspace{-.08cm}\del} \kappa^2 \Psi^{1}(\eta,\vec{x};
\vec{k},s) \longrightarrow \f{\kappa^2 H^2}{2^6\pi^2} \f{17}{2} i H a \g^0
\Psi^0(\eta,\vec{x};\vec{k},s) \,\, .
\end{eqnarray}
The only way for the left hand side to reproduce such rapid growth
is for the time derivative in
$i\hspace{-.1cm}\not{\hspace{-.08cm}\del}$ to act on a factor of
$\ln(a)$,
\begin{equation}
i\gamma^{\mu}\partial_{\mu}\ln(a)
=i\gamma^{\mu}\frac{Ha^2}{a}\delta^0_{\mu}=iHa\gamma^0\;.
\end{equation}
We can therefore write the late time limit of the
tree plus one loop mode functions as,
\begin{equation}
\Psi^{0}(\eta,\vec{x};\vec{k},s) + \kappa^2 \Psi^{1}(\eta,\vec{x};\vec{k},s)
\longrightarrow \Biggl\{1 \!+\! \f{\kappa^2 H^2}{2^6 \pi^2} \f{17}{2} \ln(a)
\Biggr\} \Psi^0(\eta,\vec{x};\vec{k},s) \,\, . \label{modefun}
\end{equation}
All other corrections actually fall off at late times. For example, those
from the $\ln(a)$ terms in (\ref{localtotal}) go like $\ln(a)/a$.

There is a clear physical interpretation for the sort of solution we see
in (\ref{modefun}). When the corrected field goes to the free field times
a constant, that constant represents a field strength renormalization. When
the quantum corrected field goes to the free field times a function of time
that is independent of the form of the free field solution, it is natural to
think in terms of a {\it time dependent field strength renormalization},
\begin{equation}
\Psi(\eta,\vec{x};\vec{k},s) \longrightarrow \frac{\Psi^0(\eta,\vec{x};
\vec{k},s)}{\sqrt{Z_2(t)}} \quad {\rm where} \quad Z_2(t) = 1 \!-\! \frac{17 
\kappa^2 H^2}{2^6 \pi^2} \ln(a) \!+\! O(\kappa^4) \; .\label{fieldrenor}
\end{equation}
Of course we only have the order $\kappa^2$ correction, so one does not know
if this behavior persists at higher orders. If no higher loop correction
supervenes, the field would switch from positive norm to negative norm at
$\ln(a) = 2^6 \pi^2/17 \kappa^2 H^2$. In any case, it is safe to conclude that
perturbation theory must break down near this time.

\section{Hartree Approximation}

The appearance of a time-dependent field strength renormalization is such a
surprising result that it is worth noting we can understand it on a
simple, qualitative level using the Hartree, or mean-field, approximation.
This technique has proved useful in a wide variety of problems from atomic
physics \cite{DRH} and statistical mechanics \cite{RKP}, to nuclear physics
\cite{HoS} and quantum field theory \cite{HJS}. Of particular relevance to our
work is the insight the Hartree approximation provides into the generation of
photon mass by inflationary particle production in SQED \cite{DDPT,DPTD,PW2}.

The idea is that we can approximate the dynamics of Fermi fields interacting
with the graviton field operator, $h_{\mu\nu}$, by taking the expectation
value of the Dirac Lagrangian in the graviton vacuum. To the order we shall
need it, the Dirac Lagrangian is \cite{MW1},
\begin{eqnarray}
\lefteqn{\mathcal{L}_{\rm Dirac} = \overline{\Psi} i\hspace{-.1cm}\not{
\hspace{-.08cm} \partial} \Psi + \frac{\kappa}2 \Bigl\{h \overline{\Psi}
i\hspace{-.1cm} \not{\hspace{-.08cm} \partial} \Psi \!-\! h^{\mu\nu}
\overline{\Psi} \gamma_{\mu} i \partial_{\nu} \Psi \!-\! h_{\mu\rho , \sigma}
\overline{\Psi} \gamma^{\mu} J^{\rho\sigma} \Psi\Bigr\} } \nonumber \\
& & + \kappa^2 \Bigl[\frac18 h^2 \!-\! \frac14 h^{\rho\sigma} h_{\rho\sigma}
\Bigr] \overline{\Psi} i\hspace{-.1cm} \not{\hspace{-.08cm} \partial} \Psi
+ \kappa^2 \Bigl[-\frac14 h h ^{\mu\nu} \!+\! \frac38 h^{\mu\rho} h_{\rho}^{~\nu}
\Bigr] \overline{\Psi} \gamma_{\mu} i \partial_{\nu} \Psi \nonumber \\
& & \hspace{-.5cm} + \kappa^2 \Bigl[-\frac14 h h_{\mu\rho , \sigma}
\!+\! \frac18 h^{\nu}_{~\rho} h_{\nu \sigma , \mu} \!+\! \frac14
(h^{\nu}_{~\mu} h_{\nu \rho})_{,\sigma} \!+\! \frac14
h^{\nu}_{~\sigma} h_{\mu\rho , \nu}\Bigr] \overline{\Psi}
\gamma^{\mu} J^{\rho\sigma} \Psi + O(\kappa^3) .
\qquad\label{DiracL}
\end{eqnarray}
Of course the expectation value of a single graviton field is zero, but the
expectation value of the product of two fields is the graviton propagator
\cite{TW2,RPW3},
\begin{eqnarray}
\lefteqn{\langle\Om\mid
T\Bigl[ h_{\mu\nu}(x) h_{\rho\sigma}(x') \Bigr] \mid\Om\rangle }
\nonumber \\
& & \hspace{1cm} =
i\D_{A}(x;x')\Bigl[\mbox{}_{\mu\nu}T^{A}_{\rho\sigma}\Bigr]+
i\D_{B}(x;x')\Bigl[\mbox{}_{\mu\nu}T^{B}_{\rho\sigma}\Bigr]+
i\D_{C}(x;x')\Bigl[\mbox{}_{\mu\nu}T^{C}_{\rho\sigma}\Bigr]
\,\, . \qquad \label{gprop}
\end{eqnarray}
The various tensor structures are,
\begin{eqnarray}
\lefteqn{\Bigl[\mbox{}_{\mu\nu}T^{A}_{\rho\sigma}\Bigr] =
2\bar{\e}_{\mu(\rho}\bar{\e}_{\sigma)\nu} - \frac{2}{D \!-\! 3}
\bar{\e}_{\mu\nu}\bar{\e}_{\rho\sigma} \quad , \quad
\Bigl[\mbox{}_{\mu\nu}T^{B}_{\rho\sigma}\Bigr] =
-4\d^0\mbox{}_{(\mu}\bar{\e}_{\nu)}\mbox{}_{(\rho}\d^0_{\sigma)}
\,\, , } \\
& & \Bigl[\mbox{}_{\mu\nu}T^{C}_{\rho\sigma}\Bigr] =
\frac2{(D \!-\! 2) (D \!-\! 3)} \Bigl[(D \!-\! 3) \d^{0}_{\mu}\d^{0}_{\nu}
+ \bar{\e}_{\mu\nu}\Bigr] \Bigl[(D \!-\! 3) \d^{0}_{\rho}\d^{0}_{\sigma}
+ \bar{\e}_{\rho\sigma}\Bigr] \,\, . \qquad \label{gtensor}
\end{eqnarray}
Parenthesized indices are symmetrized and a bar over a common tensor such
as the Kronecker delta function denotes that its temporal components
have been nulled,
\begin{equation}
\overline{\delta}^{\mu}_{\nu} \equiv \delta^{\mu}_{\nu} -
\delta^{\mu}_{\scriptscriptstyle 0} \delta^0_{\nu} \qquad , \qquad
\overline{\eta}_{\mu\nu} \equiv \eta_{\mu\nu} + \delta^0_{\mu}
\delta^0_{\nu} \; .
\end{equation}

The three scalar propagators that appear in (\ref{gprop}) have complicated
expressions which we omit in favor of simply giving their coincidence limits
and the coincidence limits of their first derivatives \cite{TW3},
\begin{eqnarray}
\lim_{x' \rightarrow x} \, {i\Delta}_A(x;x') & = & \frac{H^{D-2}}{(4\pi)^{
\frac{D}2}} \frac{\Gamma(D-1)}{\Gamma(\frac{D}2)} \left\{-\pi \cot\Bigl(
\frac{\pi}2 D \Bigr) + 2 \ln(a) \right\} , \\
\lim_{x' \rightarrow x} \, \partial_{\mu} {i\Delta}_A(x;x') & = &
\frac{H^{D-2}}{(4\pi)^{\frac{D}2}} \frac{\Gamma(D-1)}{\Gamma(\frac{D}2)}
\times H a \delta^0_{\mu} = \lim_{x' \rightarrow x} \, \partial_{\mu}'
{i\Delta}_A(x;x') \; , \qquad \\
\lim_{x' \rightarrow x} \, {i\Delta}_B(x;x') & = & \frac{H^{D-2}}{(4\pi)^{
\frac{D}2}} \frac{\Gamma(D-1)}{\Gamma(\frac{D}2)}\times -\frac1{D\!-\!2} \; ,\\
\lim_{x' \rightarrow x} \, \partial_{\mu} {i\Delta}_B(x;x') & = & 0 =
\lim_{x' \rightarrow x} \, \partial_{\mu}' {i\Delta}_B(x;x') \; , \\
\lim_{x' \rightarrow x} \, {i\Delta}_C(x;x') & = & \frac{H^{D-2}}{(4\pi)^{
\frac{D}2}} \frac{\Gamma(D-1)}{\Gamma(\frac{D}2)}\times \frac1{(D\!-\!2)
(D\!-\!3)} \; ,\\
\lim_{x' \rightarrow x} \, \partial_{\mu} {i\Delta}_C(x;x') & = & 0 =
\lim_{x' \rightarrow x} \, \partial_{\mu}' {i\Delta}_C(x;x') \; .
\end{eqnarray}
We are interested in terms which grow at late times. Because the $B$-type
and $C$-type propagators go to constants, and their derivatives vanish, they
can be neglected. The same is true for the divergent constant in the
coincidence limit of the $A$-type propagator. In the full theory it would be
absorbed into a constant counterterm. Because the remaining, time dependent
terms are finite, we may as well take $D \!=\! 4$. Our Hartree approximation
therefore amounts to making the following replacements in (\ref{DiracL}),
\begin{eqnarray}
h_{\mu\nu} h_{\rho\sigma} & \longrightarrow & \frac{H^2}{4 \pi^2} \ln(a)
\Bigl[ \overline{\eta}_{\mu\rho} \overline{\eta}_{ \nu\sigma} \!+\!
\overline{\eta}_{\mu\sigma} \overline{\eta}_{\nu\rho} \!-\! 2
\overline{\eta}_{\mu\nu} \overline{\eta}_{\rho\sigma} \Bigr] \; ,
\label{Atensor1} \\
h_{\mu\nu} h_{\rho\sigma, \alpha} & \longrightarrow &
\frac{H^2}{8 \pi^2} H a \delta^0_{\alpha} \, \Bigl[\overline{\eta}_{\mu\rho}
\overline{\eta}_{\nu\sigma} \!+\! \overline{\eta}_{\mu\sigma}
\overline{\eta}_{\nu\rho} \!-\! 2 \overline{\eta}_{\mu\nu}
\overline{\eta}_{\rho\sigma} \Bigr] \; . \label{Atensor2}
\end{eqnarray}

It is now just a matter of contracting (\ref{Atensor1}-\ref{Atensor2})
appropriately to produce each of the quadratic terms in (\ref{DiracL}).
For example, the first term gives,
\begin{eqnarray}
\f{\kappa^2}{8} h^2 \overline{\Psi} i \hspace{-.1cm} \not{
\hspace{-.08cm}\del} \Psi &\!\!\!\! \longrightarrow \!\!\!\!&
\f{\kappa^2 H^2}{2^5\pi^2}
\ln(a) \Bigl[\e^{\mu\nu}\e^{\rho\sigma}\Bigr]\Bigl[\bar{\e}_{\mu\rho}
\bar{\e}_{\nu\sigma} +\bar{\e}_{\mu\sigma}\bar{\e}_{\nu\rho}
-2\bar{\e}_{\mu\nu}\bar{\e}_{\rho\sigma}\Bigr]
\overline{\Psi}i\hspace{-.1cm}\not{\hspace{-.08cm}\del}\Psi ,
\qquad \\
& \!\!\!\! = \!\!\!\! &
\f{\kappa^2 H^2}{2^5\pi^2}\ln(a)\Bigl[3+3-18\Bigr]
\overline{\Psi}i\hspace{-.1cm}\not{\hspace{-.08cm}\del}\Psi \,\,.
\end{eqnarray}
The second quadratic term gives a proportional result,
\begin{eqnarray}
\f{-\kappa^2}{4}h^{\rho\sigma}h_{\rho\sigma}
\overline{\Psi}i\hspace{-.1cm}\not{\hspace{-.08cm}\del}\Psi
\longrightarrow\f{-\kappa^2 H^2}{2^4\pi^2}\ln(a)\bigl[9+3-6\Bigr]
\overline{\Psi}i\hspace{-.1cm}\not{\hspace{-.08cm}\del}\Psi \,\, .
\end{eqnarray}
The total for these first two terms is $\f{-3\kappa^2 H^2}{4\pi^2} \ln(a)
\overline{\Psi}i\hspace{-.1cm} \not{\hspace{-.08cm}\del}\Psi$.

The third and fourth of the quadratic terms in (\ref{DiracL}) result in
only spatial derivatives,
\begin{eqnarray}
\f{-\kappa^2H^2}{4}hh^{\mu\nu}\overline{\Psi}\g_{\mu}i\del_{\nu}\Psi
& \longrightarrow & \f{-\kappa^2 H^2}{2^4\pi^2}\ln(a)\Bigl[1+1-6\Bigr]
\overline{\Psi}i\hspace{-.1cm}\not{\hspace{-.08cm}\bar{\del}}\Psi
\,\, , \\
\f{3}{8}\kappa^2h^{\mu\rho}h_{\,\rho}^{\nu}
\overline{\Psi}\g_{\mu}i\del_{\nu}\Psi & \longrightarrow &
\f{3\kappa^2H^2}{2^5\pi^2}\ln(a)\Bigl[3+1-2\Bigr]
\overline{\Psi}i\hspace{-.1cm}\not{\hspace{-.08cm}\bar{\del}}\Psi \,\, .
\end{eqnarray}
The total for this type of contribution is $\f{7\kappa^2 H^2}{2^4\pi^2}
\ln(a) \overline{\Psi} i \hspace{-.1cm} \not{\hspace{-.08cm} \bar{\del}}
\Psi$.

The final four quadratic terms in (\ref{DiracL}) involve derivatives acting
on at least one of the two graviton fields,
\begin{eqnarray}
-\f{\kappa^2}{4}hh_{\mu\rho,\si}\overline{\Psi}\g^{\mu}J^{\rho\si}\Psi
& \longrightarrow & \f{-\kappa^2 H^2}{2^5\pi^2}Ha\Bigl[1+1-6\Bigr]
\bar{\e}_{\mu\rho}\overline{\Psi}\g^{\mu}J^{\rho0}\Psi \; , \\
\f{\kappa^2}{8}h^{\nu}_{\,\rho}h_{\nu\si,\mu}
\overline{\Psi}\g^{\mu}J^{\rho\si}\Psi & \longrightarrow &
\f{\kappa^2H^2}{2^6\pi^2}Ha\Bigl[3+1-2\Bigr] \bar{\e}_{\rho\si}
\overline{\Psi}\g^0J^{\rho\si}\Psi \; , \\
\f{\kappa^2}{4}\Bigl(h^{\nu}_{\,\mu}h_{\nu\rho}\Bigr)_{,\sigma}
\overline{\Psi}\g^{\mu}J^{\rho\si}\Psi & \longrightarrow &
\f{\kappa^2H^2}{2^4\pi^2}Ha\Bigl[3+1-2\Bigr]\e_{\mu\rho}
\overline{\Psi}\g^{\mu}J^{\rho0}\Psi \; , \\
\f{\kappa^2}{4}h^{\nu}_{\,\si}h_{\mu\rho,\nu}
\overline{\Psi}\g^{\mu}J^{\rho\si}\Psi & \longrightarrow & 0 \; .
\end{eqnarray}
The second of these contributions vanishes owing to the anti\-sym\-met\-ry
of the Lorentz representation matrices, $J^{\mu\nu} \equiv \frac{i}4
[\gamma^{\mu},\gamma^{\nu}]$, whereas $\overline{\eta}_{\mu \rho}
\gamma^{\mu} J^{\rho 0} = -\frac{3i}2 \gamma^0$. Hence the sum of all
four terms is $\f{-3\kappa^2H^2}{8\pi^2} H a \overline{\Psi} i \g^0 \Psi$.

Combining these results gives,
\begin{eqnarray}
\lefteqn{\Bigl\langle \mathcal{L}_{\rm Dirac} \Bigr\rangle =
\overline{\Psi} i\!\hspace{-.1cm}\not{\hspace{-.08cm} \partial} \Psi
-\frac{3 \kappa^2 H^2}{4 \pi^2} \ln(a) \overline{\Psi}
i\hspace{-.1cm} \not{\hspace{-.08cm} \partial}
\Psi } \nonumber \\
& & \hspace{3cm} - \frac{3 \kappa^2 H^2}{8 \pi^2} H a \overline{\Psi} i\gamma^0
\Psi \!+\! \frac{7 \kappa^2 H^2}{16 \pi^2} \ln(a) \overline{\Psi} i \,
\hspace{-.1cm} \overline{\not{\hspace{-.08cm} \partial}} \Psi \!+\! O(\kappa^4)
, \\
& & \hspace{-.7cm} = \!\overline{\Psi} \Bigl[1 \!-\! \frac{3
\kappa^2 H^2}{8 \pi^2} \ln(a)\Bigr]
i\!\hspace{-.1cm}\not{\hspace{-.08cm} \partial} \Bigl[1 \!-\!
\frac{3 \kappa^2 H^2}{8 \pi^2} \ln(a)\Bigr] \Psi \!+\! \frac{7
\kappa^2 H^2}{16 \pi^2} \ln(a) \overline{\Psi} i \, \hspace{-.1cm}
\overline{\not{ \hspace{-.08cm} \partial}} \Psi \!+\! O(\kappa^4) .
\qquad\label{Hartreesum}
\end{eqnarray}
If we express the equations associated with (\ref{Hartreesum}) according to
the perturbative scheme of Section 1, the first order equation is,
\begin{eqnarray}
i\hspace{-.1cm}\not{\hspace{-.08cm}\del} \kappa^2 \Psi^{1}(\eta,\vec{x};
\vec{k},s) = \f{\kappa^2 H^2}{2^6\pi^2} i H \g^0 \Psi^0(\eta,\vec{x};\vec{k},s)
\Bigl\{ 24 a - 28 i \frac{k}{H}\ln(a) \Bigr\} \,\, . \label{Hareqn}
\end{eqnarray}
This is similar, but not identical to, what we got in expression
(\ref{localtotal}) from the delta function terms of the actual one loop
self-energy (\ref{ren}). In particular, the exact calculation gives
$\frac{17}{2} a
\!-\! 14 i \frac{k}H \ln(a)$, rather than the Hartree approximation
of $24 a \!-\! 28 i \frac{k}H \ln(a)$. Of course the $\ln(a)$ terms
make corrections to $\Psi^1$ which fall like $\ln(a)/a$, so the real
disagreement between the two methods is limited to the differing factors
of $\frac{17}{2}$ versus $24$.

We are pleased that such a simple technique comes so close to recovering
the result of a long and tedious calculation. The slight discrepancy is
no doubt due to terms in the Dirac Lagrangian (\ref{DiracL}) which are
linear in the graviton field operator. As described in relation (\ref{SKop})
of section 2, the linearized effective field $\Psi_i(x;\vec{k},s)$ represents 
$a^{\frac{D-1}2}$ times the expectation value of the anti-commutator of the 
Heisenberg field operator $\psi_i(x)$ with the free fermion creation operator
$b(\vec{k},s)$. At the order we are working, quantum corrections to 
$\Psi_i(x;\vec{k},s)$ derive from perturbative corrections to $\psi_i(x)$ 
which are quadratic in the free graviton creation and annihilation operators.
Some of these corrections come from a single $h h \overline{\psi} \psi$ vertex,
while others derive from two $h \overline{\psi} \psi$ vertices. The Hartree
approximation recovers corrections of the first kind, but not the second, 
which is why we believe it fails to agree with the exact result.
Yukawa theory presents a fully worked-out example \cite{PW1,GP,MW2}
in which the {\it entire} lowest-order correction to the fermion mode
functions derives from the product of two such linear terms, so the Hartree
approximation fails completely in that case.

\section{Discussion}

We have used the Schwinger-Keldysh formalism to include one loop, quantum
gravitational corrections to the Dirac equation, in the simplest local Lorentz
and general coordinate gauge, in the locally de Sitter
background which is a paradigm for inflation. Because Dirac + Einstein is
not perturbatively renormalizable, it makes no sense to solve this equation
generally. However, the equation should give reliable predictions at late
times when the arbitrary finite parts of the BPHZ counterterms (\ref{genctm})
are insignificant compared to the completely determined factors of
$\ln(a a')$ on terms (\ref{1stlog}-\ref{3rdlog}) which otherwise have the
same structure. In this late time limit we find that the one loop corrected,
spatial plane wave mode functions behave as if the tree order mode functions
were simply subject to a time-dependent field strength renormalization,
\begin{eqnarray}
Z_2(t) = 1 - \f{17}{4\pi} G H^2 \ln(a) + O(G^2) \,\,\,\,{\rm where}\,\,\,\,
G=16\pi\kappa^2\,\,.
\end{eqnarray}
If unchecked by higher loop effects, this would vanish at $\ln(a) \simeq
1/G H^2$. What actually happens depends upon higher order corrections, but
there is no way to avoid perturbation theory breaking down at this time,
at least in this gauge.

Might this result be a gauge artifact? One reaches different gauges by
making field dependent transformations of the Heisenberg operators. We have 
worked out the change (\ref{SKprime}) this induces in the linearized 
effective field, but the result is not simple. Although the linearized 
effective field obviously changes when different gauge conditions are
employed to compute it, we believe (but have not proven) that the late time 
factors of $\ln(a)$ do not change.

It is important to realize that the 1PI functions of a gauge theory in a
fixed gauge are not devoid of physical content by virtue of depending upon
the gauge. In fact, they encapsulate the physics of a quantum gauge field
every bit as completely as they do when no gauge symmetry is present. One 
extracts this physics by forming the 1PI functions into gauge independent 
and physically meaningful combinations. The S-matrix accomplishes this in 
flat space quantum field theory. Unfortunately, the S-matrix fails to exist 
for Dirac + Einstein in de Sitter background, nor would it correspond to 
an experiment that could be performed if it did exist \cite{TW8,EW,AS}. 

If it is conceded that we know what it means to release the universe in a
free state then it would be simple enough --- albeit tedious --- to construct
an analogue of $\psi_i(x)$ which is invariant under gauge transformations
that do not affect the initial value surface. For example, one might extend
to fermions the treatment given for pure gravity by \cite{TW9}:
\begin{itemize}
\item{Propagate an operator-valued geodesic a fixed invariant time from the 
initial value surface;}
\item{Use the spin connection $A_{\mu cd} J^{cd}$ to parallel transport along
the ge\-o\-des\-ic; and}
\item{Evaluate $\psi$ at the operator-valued geodesic, in the Lorentz frame
which is transported from the initial value surface.}
\end{itemize}
This would make an invariant, as would any number of other constructions
\cite{GMH}. For that matter, the gauge-fixed 1PI functions also correspond 
to the expectation values of invariant operators \cite{RPW5}. Mere invariance 
does not guarantee physical significance, nor does gauge dependence preclude 
it. 

What is needed is for the community to agree upon a relatively simple set of 
operators which stand for experiments that could be performed in de Sitter 
space. There is every reason to expect a successful outcome because the last 
few years have witnessed a resolution of the similar issue of how to measure 
quantum gravitational back-reaction during inflation, driven either by a 
scalar inflaton \cite{WU,AW1,AW2,GB} or by a bare cosmological constant 
\cite{TW10}. That process has begun for quantum field theory in de Sitter
space \cite{EW,AS,GMH} and one must wait for it to run its course. In the
meantime, it is safest to stick with what we have actually shown: perturbation
theory must break down for Dirac + Einstein in the simplest gauge.

This is a surprising result but we were able to understand it qualitatively
using the Hartree approximation in which one takes the expectation value of
the Dirac Lagrangian in the graviton vacuum. The physical interpretation
seems to be that fermions propagate through an effective geometry whose
ever-increasing deviation from de Sitter is controlled by inflationary
graviton production. At one loop order the fermions are passive spectators
to this effective geometry.

It is significant that inflationary graviton production enhances fermion
mode functions by a factor of $\ln(a)$ at one loop. Similar factors of
$\ln(a)$ have been found in the graviton vacuum energy \cite{TW4,TW5}.
These infrared logarithms also occur in the vacuum energy of a massless,
minimally coupled scalar with a quartic self-interaction \cite{OW1,OW2},
and in the VEV's of almost all operators in Yukawa theory \cite{MW2}
and SQED \cite{PTW}. A recent all orders analysis was not even able to
exclude the possibility that they might contaminate the power spectrum
of primordial density fluctuations \cite{SW2}!

The fact that infrared logarithms grow without bound raises the
exciting possibility that quantum gravitational corrections may be
significant during inflation, in spite of the minuscule coupling
constant of $G H^2 \ltwid 10^{-12}$. However, the only thing one can
legitimately conclude from the perturbative analysis is that
infrared logarithms cause perturbation theory to break down, in our
gauge, if inflation lasts long enough. Inferring what happens after 
this breakdown requires a nonperturbative technique.

Starobinski\u{\i} has long advocated that a simple stochastic formulation
of scalar potential models serves to reproduce the leading infrared
logarithms of these models at each order in perturbation theory \cite{AAS}.
This fact has recently been proved to all orders \cite{RPW4,TW6}. When the
scalar potential is bounded below it is even possible to sum the series of
leading infrared logarithms and infer their net effect at asymptotically
late times \cite{SY}! Applying Starobinski\u{\i}'s technique to more
complicated theories which also show infrared logarithms is a formidable
problem, but solutions have recently been obtained for Yukawa theory
\cite{MW2} and for SQED \cite{PTW}. It would be very interesting to see
what this technique gives for the infrared logarithms we have exhibited,
to lowest order, in Dirac + Einstein. And it should be noted that even the
potentially complicated, invariant operators which might be required to
settle the gauge issue would be straightforward to compute in such a 
stochastic formulation.

\section{Appendix: nonlocal terms from section 4}

It is important to establish that the nonlocal terms make no
significant contribution at late times, so we will derive the results
summarized in Table \ref{noncon}. For simplicity we denote as $[U^I]$
the contribution from each operator $U^I_{ij}$ in Table \ref{nond}.
We also abbreviate $\Psi^0(\e,\vec{x};\vec{k},s)$ as $\Psi^0(x)$.

Owing to the factor of $1/a'$ in $U^1_{ij}$, and to the larger number
of derivatives, the reduction of $[U^1]$ is atypical,
\begin{eqnarray}
\lefteqn{[{\rm U^1}] \equiv \f{\kappa^2}{2^8\pi^4} \f{1}{a}
\hspace{-.1cm} \not{\hspace{-.08cm}\del}\del^4\int d^4x'\f{1}{a'}
\Biggr\{\f{\ln(\mu^2\D x^2_{\scriptscriptstyle ++})}{\D x^2_{
\scriptscriptstyle ++}}-\f{\ln(\mu^2\D x^2_{\scriptscriptstyle +-})}{
\D x^2_{\scriptscriptstyle +-}}\Biggl\}\Psi^0(x') \; , } \\
&&=\f{-i\kappa^2}{2^6\pi^2a}\g^0\Psi^0(x)\Bigl[-2ik\del_{\e}
+\del_{\e}^{2}\Bigr]\Biggl\{\del_{\e}\int_{\e_i}^{\e} \!\! d\eta'
(-H\e') \Bigl(\f{e^{2ik\D\e}-1}{\Delta \eta}\Bigr) \nonumber \\
&& \hspace{6cm} + \del_{\e}^2 \int_{\e_i}^{\e} \!\! d\eta'
(-2H\e')\ln(2\mu\D\e)\Biggr\} \; , \\
&& = \f{-i\kappa^2}{2^6\pi^2a}\g^0\Psi^0\Bigl(-2ik+\del_{\e}\Bigr)
\Biggl\{-\f{e^{2ik(\e+\f{1}{H})}-1}{(\e+\f{1}{H})^2}\nn\\
&& \hspace{2cm} +\f{(2ik-H)e^{2ik(\e+\f{1}{H})}}{\e+\f{1}{H}}
-\f{3H^2}{(1+H\e)}+\f{2H^3\e}{(1+H\e)^2}\Biggr\}\; , \\
&& = \f{\kappa^2H^2}{2^6\pi^2}(H\e)iH\g^0\Psi\Biggl\{\f{2\Bigl[
e^{\f{2ik}{H}(1+H\e)}-1-2H\e\Bigr]}{(1+H\e)^3}+\f{(1-\f{2ik}{H})
e^{\f{2ik}{H}(1+H\e)}}{(1+H\e)^2}\nn\\
&& \hspace{6cm} + \f{5-4ik\e-\f{2ik}{H}}{(1+H\e)^2}+
\f{\f{6ik}{H}}{1+H\e}\Biggr\} \,\, .
\end{eqnarray}
This expression actually vanishes in the late time limit of $\eta \!
\rightarrow \! 0^-$.

$[U^2]$ was reduced in Section 4 so we continue with $[U^3]$,
\begin{eqnarray}
\lefteqn{[U^3] \equiv -\f{\kappa^2H^2}{2^8\pi^4} \hspace{-.1cm}
\not{\hspace{-.08cm}\bar{\del}} \del^2 \!\! \int \!  d^4x'\Biggr\{
\f{\ln(\mu^2\D x^2_{\scriptscriptstyle ++})}{\D x^2_{\scriptscriptstyle
++}} - \f{\ln(\mu^2\D x^2_{\scriptscriptstyle +-})}{\D x^2_{
\scriptscriptstyle +-}} \Biggl\} \Psi^0(x') \; , } \\
&& = -\f{\kappa^2H^2}{2^8\pi^4}\hspace{-.1cm}
\not{\hspace{-.08cm}\bar{\del}}i4\pi^2\Psi^0(x)\Biggl\{
2\ln\Bigl[\f{2\mu}{H}(1+H\e)\Bigr] + \int_{\e_i}^{\e} d\e'
\Bigl(\f{e^{2ik\D\e}-1}{\D\e}\Bigr)\Biggr\} \; , \\
&& = \f{\kappa^2H^2}{2^6\pi^2}k\g^0\Psi^0(x)\Biggl\{2\ln\Bigl[
\f{2\mu}{H}(1+H\e)\Bigr]+\int_{\e_i}^{\e}d\e'
\Bigl(\f{e^{2ik\D\e}-1}{\D\e}\Bigr)\Biggr\} \; , \\
&& \longrightarrow \f{\kappa^2H^2}{2^6\pi^2}iH\g^0\Psi^0(x) \times
-\f{i k}{H} \Biggl\{2\ln(\f{2\mu}{H})-\int_{\e_i}^{0}d\e'
\Bigl(\f{e^{-2ik\e'}-1}{\e'}\Bigr)\Biggr\} . \qquad \label{U3}
\end{eqnarray}
$U^4_{ij}$ has the same derivative structure as $U^3_{ij}$, so $[U^4]$
follows from (\ref{U3}),
\begin{eqnarray}
\lefteqn{[U^4] \equiv -\f{\kappa^2H^2}{2^8\pi^4} \times 8 \hspace{-.1cm}
\not{\hspace{-.08cm}\bar{\del}} \del^2 \!\! \int \! d^4x'\Biggr\{
\f{\ln(\frac14 H^2 \D x^2_{\scriptscriptstyle ++})}{\D x^2_{
\scriptscriptstyle ++}} - \f{\ln(\frac14 H^2 \D x^2_{\scriptscriptstyle
+-})}{\D x^2_{\scriptscriptstyle +-}} \Biggl\} \Psi^0(x') \; , } \\
&& \hspace{1.3cm} = \f{\kappa^2H^2}{2^6\pi^2}8k\g^0\Psi^0(x)\Biggl\{2
\ln\Bigl[ (1+H\e)\Bigr]+\int_{\e_i}^{\e}d\e'
\Bigl(\f{e^{2ik\D\e}-1}{\D\e}\Bigr)\Biggr\} \; , \qquad \\
&& \hspace{1.3cm} \longrightarrow \f{\kappa^2H^2}{2^6\pi^2} i H \g^0
\Psi^0(x) \times 8 i \f{k}{H} \int_{\e_i}^{0} \!\! d\e'
\Bigl(\f{e^{-2ik\e'}-1}{\e'}\Bigr) \; .
\end{eqnarray}

$U^5_{ij}$ has a Laplacian rather than a d'Alembertian so we use
identity (\ref{key1}) for $[U^5]$. We also employ the abbreviation
$k\D\e \!=\! \a$,
\begin{eqnarray}
\lefteqn{[U^5] \equiv 4 \f{\kappa^2H^2}{2^8\pi^4} \hspace{-.1cm}
\not{\hspace{-.08cm} \del} \nabla^2 \!\! \int \! d^4x'\Biggr\{\f{
\ln(\mu^2\D x^2_{\scriptscriptstyle ++})}{\D x^2_{\scriptscriptstyle
++}} - \f{\ln(\mu^2\D x^2_{\scriptscriptstyle +-})}{\D x^2_{
\scriptscriptstyle +-}} \Biggl\} \Psi^0(x') \; , } \\
&& \hspace{-.5cm} = 4\f{\kappa^2H^2}{2^8\pi^4}\hspace{-.1cm}
\not{\hspace{-.08cm}\del} \nabla^2 \Bigl( \f{-4i\pi^2}{k} \Bigr)
\Psi^0(x) \int_{\e_i}^{\e}d\e' e^{i\a} \nn \\
&& \hspace{0cm} \times \Biggl\{\!-\cos(\a) \! \int_{0}^{2\a} \!\!\!
dt \, \f{\sin(t)}{t} + \sin(\a) \Bigl[\int_{0}^{2\a} \!\!\! dt \Bigl(
\f{\cos(t) \!-\! 1}{t}\Bigr) + 2\ln\Bigl(\f{H\a}{k}\Bigr)\Bigr] \!
\Biggr\} , \qquad \\
&& \hspace{-.5cm} = \f{\kappa^2H^2}{2^6\pi^2} i H \g^0 \Psi^0(x)
\times 4\f{k^2}{H} \int_{\e_i}^{\e} \!\! d\e'e^{2i\a}\Bigl[\int_{0}^{2\a}
\!\!\! dt\Bigl( \f{e^{-it}-1}{t}\Bigr)+\ln(H\D\e)^2\Bigr] \; , \qquad \\
&& \hspace{-.5cm} \longrightarrow \f{\kappa^2H^2}{2^6\pi^2}iH\g^0
\Psi^0(x) \times 4 \f{k^2}{H} \int_{\e_i}^{0} \!\! d\e'e^{2i\a} \Bigl[
\int_{0}^{2\a} \!\!\! dt \Bigl(\f{e^{-it}-1}{t}\Bigr)+\ln(H\e')^2\Bigr] .
\end{eqnarray}
$U^6_{ij}$ has the same derivative structure as $U^5_{ij}$ but it acts
on a different integrand. We therefore apply identity (\ref{key3}) for
$[U^6]$,
\begin{eqnarray}
\lefteqn{[U^6] \equiv 7 \f{\kappa^2H^2}{2^8\pi^4} \hspace{-.1cm}
\not{\hspace{-.08cm}\del} \nabla^2 \!\! \int \! d^4x' \Biggr\{
\f{1}{\D x^2_{++}}-\f{1}{\D x^2_{+-}}\Biggl\}\Psi^0(x') \; , } \\
&& = 7\f{\kappa^2H^2}{2^8\pi^4}\hspace{-.1cm}
\not{\hspace{-.08cm}\del}\nabla^2\times(-i4\pi^2)k^{-1}
\Psi^0(x)\int_{\e_i}^{\e}d\e'e^{ik\D\e}\sin(k\D\e) \; , \\
&&= \f{\kappa^2H^2}{2^6\pi^2}iH\g^0\Psi^0(x)\times-\f{7}{2}\f{ik}{H}
\Bigl[e^{\f{2ik}{H}(1+H\e)}-1\Bigr] \; , \\
&&\longrightarrow\f{\kappa^2H^2}{2^6\pi^2}iH\g^0\Psi^0(x)
\times-\f{7}{2}\f{ik}{H} \Bigl[e^{\f{2ik}{H}}-1\Bigr] \; .
\end{eqnarray}

\vskip .3cm

\centerline{\bf Acknowledgements}

This work was partially supported by NSF grant PHY-0244714 and by
the Institute for Fundamental Theory at the University of Florida.

\end{document}